\documentclass{aa} 
\usepackage{hyperref}
\usepackage{lscape}

\usepackage{csvsimple}
\usepackage{amsmath}	
\usepackage{booktabs}
\usepackage{array}
\usepackage{longtable}
\usepackage[flushleft]{threeparttable}
\usepackage{soul}
\usepackage[switch]{lineno}

\usepackage{graphicx}

\makeatletter
\newcommand{\@ssymbol}[1]{\ifcase#1\or\alpha\or\beta\or\gamma\or\delta\or\epsilon\or\varepsilon
  \or\zeta\or\eta\or\theta\or\vartheta\or\iota\or\kappa\or\lambda\or\mu\or\nu\or\xi\or\pi
  \or\varpi\or\rho\or\varrho\or\sigma\or\varsigma\or\tau\or\upsilon\or\phi\or\varphi\or\chi
  \or\psi\or\omega\else\@ctrerr\fi}
\newcommand{\ssymbol}[1]{^{\@ssymbol{#1}}}
\makeatother

\usepackage{txfonts}
\begin{document}

   \title{Detection of a new sample of Galactic White Dwarfs in the direction of the Small Magellanic Cloud}

   \subtitle{}
   
   \author{A.V. Sidharth$^{1}$\thanks{E-mail: sidharthnarayanan17@gmail.com},
B. Shridharan$^{1,3}$, Blesson Mathew$^1$, A. Devaraj$^1$, T.B. Cysil$^1$, C. S. Stalin$^2$,
R. Arun$^2$, Suman Bhattacharyya$^1$, Sreeja S. Kartha$^1$, and T. Robin$^1$}
    \authorrunning{Sidharth et al.}
   \institute{Department of Physics and Electronics, CHRIST (Deemed to be University), Bangalore 560029, India
   \and Indian Institute of Astrophysics, Sarjapur Road, Koramangala, Bangalore 560034, India
   \and Tata Institute of Fundamental Research, Homi Bhabha Road, Mumbai 400005, India}

   \date{Received xxxx; accepted xxxx}

 
  \abstract
   {}
   {In this study, we demonstrate the efficacy of the Ultraviolet Imaging Telescope (UVIT) in identifying and characterizing white dwarfs (WDs) within the Milky Way Galaxy. }
   {Leveraging the UVIT point source catalogue towards SMC and crossmatching with Gaia DR3 data, we identified 43 single WDs (37 new detections), 13 new WD+main sequence (MS) candidates and 161 UV bright MS stars by analysing their Spectral Energy Distributions (SED). Using the WD evolutionary models, we determine the masses, effective temperatures, and cooling ages of these identified WDs.}
   { Masses of these WDs range from 0.2 to 1.3 M$_{\odot}$ and effective temperatures (T$_{eff}$) between 10000 K to 15000 K, with cooling ages spanning 0.1 to 2 Gyr. Notably, we detect hotter WDs compared to literature values, which is attributed to the sensitivity of UVIT. Furthermore, we report the detection of 20 new extremely low-mass (ELM) candidates from our analysis. Future spectroscopic studies of the ELM candidates will help us understand the formation scenarios of these exotic objects. Despite limitations in Gaia DR3 distance measurements for optically faint WDs, we provide a crude estimate of the WD space density within 1kpc as 1.3 $\times$ 10$^{-3}$ pc$^{-3}$, which is higher than previous estimates in the literature. }
  {Our results underscore the instrumental capabilities of UVIT and anticipate the forthcoming UV missions like INSIST for systematic WD discovery. Our methodology sets a precedent for future analyses in other UVIT fields to find more WDs and perform spectroscopic studies to verify their candidacy.}


   \maketitle
%

\section{Introduction}

White Dwarfs (WDs) are the end stages of stellar evolution for the majority of main-sequence (MS) stars with masses less than $8~\textup{M}_\odot$, which live out their life dissipating their remnant energy and cool down \citep{fontaine2001potential}. The census and characterization of WDs have witnessed significant advancements, propelled by the utilization of various optical surveys such as Gaia Data Release 2 (DR2) \citep{gaia2018gaia}, Gaia Data Release 3 (DR3) \citep{brown2021gaia}, and Sloan Digital Sky Survey (SDSS) \citep{york2000sloan}. Works of \citet{gentile2019Gaia, gentile2021catalogue}, \citet{kepler2016new, kepler2019white, kepler2021white}, and \citet{eisenstein2006catalog} have significantly refined our understanding of WDs. Accurate parallax measurements from Gaia DR2 and DR3 have revolutionized the study of WDs, enabling an unprecedented scale of search for these stellar remnants. \citet{jimenez2018white} identified $\sim$ 73,000 WD candidates, delving into a detailed exploration of the population within the 100 pc solar neighbourhood. Extending this effort, \citet{gentile2019Gaia} used Gaia DR2 to identify $\sim$ 260,000 WD candidates, whereas \citet{gentile2021catalogue} presented a compilation of $\sim$ 359,000 high-confidence WD candidates spanning all sky using Gaia EDR3.

On the other hand, hot WDs (with $T_{eff} \geq$ 10000 K) remained elusive in these optical surveys as they have low optical luminosity and optical colours are insensitive to hotter temperatures \citep{gomez2007uv}. The operation of the Galaxy Evolution Explorer (GALEX)  ultraviolet sky survey \citep{morrissey2007calibration}, \citet{bianchi2011catalogues}, utilized GALEX to identify hot WDs in the Milky Way. UV point source catalogues stand as invaluable gateways into the studies of UV-bright MS stars, blue straggler stars (BSS), yellow straggler stars, sub–subgiants, WDs, and white dwarf -- main sequence (WD+MS) binary systems \citep{bianchi2009ultraviolet, bianchi2011catalogues, parsons2016white, rebassa2017white, subramaniam2018ultraviolet,ren2020white}. The UV surveys, in combination with optical photometry, help us to characterize these systems and understand the end stages of stellar evolution. \citet{rebassa2021white} studied WD+MS systems using Gaia EDR3 to identify a volume-limited sample of 112 unresolved WD+MS binaries within 100 pc. Combining astrometric and photometric data from Gaia DR3 with GALEX GR6/7, \citet{nayak2022hunting} identified 93 WD+MS candidates. Recently, \citet{jackim2024galex} presented the GALEX - Gaia EDR3 catalogue containing 332,111 candidate WD binary systems and 111,996 candidate single white dwarfs.

Ultra Violet Imaging Telescope (UVIT) onboard AstroSat mission is a suite of Far Ultra Violet (FUV: 130 - 180 nm), Near Ultra Violet (NUV: 200 - 300 nm) and Visible band (VIS: 320-550 nm) imagers. UVIT can perform simultaneous observations in these three channels with a field of view (FoV) diameter of $\sim$~28\arcmin and an angular resolution of 1.5-1.8\arcsec (more details on UVIT can be found in \citealp{kumar2012ultraviolet,tandon2017orbit} and \citealp{tandon2017orbitb}). The higher angular resolution of UVIT, compared to the 4.2-5.3\arcsec angular resolution of GALEX \citep{morrissey2007calibration}, and the availability of seven filters in the UV range (1161~\AA\ -- 2882~\AA) compared to two filters in GALEX, help us characterise WD systems with better precision compared to previous studies conducted using GALEX. Previous studies using UVIT observations have identified WDs and WD binary systems in open clusters and globular clusters, as evident from a series of UVIT Open Cluster studies (eg: \citealp{panthi2024uocs,panthi2023field,vaidya2022uocs,sindhu2019uvit,jadhav2019uvit}) and Globular Cluster UVIT Legacy Survey (GlobULeS) studies (eg: \citealp{dattatrey2023globule,dattatrey2023globules,prabhu2022globular,sahu2022globular}), thus showcasing the superior capabilities of the UVIT instrument.    

In this study, we use the UVIT point-source catalogue from \citet{devaraj2023uvit} to explore the FUV(1300 -- 1800 \AA) and NUV (2000 -- 3000 \AA) observations directed towards a previously less explored line of sight towards the Small Magellanic Cloud (SMC). As a follow-up to the UVIT point-source catalogue provided by \citet{devaraj2023uvit}, we crossmatch them with the Gaia DR3 catalogue \citep{vallenari2023gaia}. This allows us to combine the photometric data from 7 UV filters and the optical photometry and astrometry from Gaia DR3 in order to identify and characterise the UV bright sources in the FoV of the SMC. The availability of 7 UV data points allows the generation of a better fit to the Spectral Energy Distributions (SED), providing more accurate parameters.  

The manuscript is structured as follows. In Section 2, we describe the data used in this study and discuss the methodology employed for cross-matching the UVIT point-source catalogue with Gaia DR3. Section 3 presents the methodology for identifying MS, WD and WD+MS systems in our sample. Further, by combining UV photometry with available optical photometry, we fit the SED of each source to characterise the identified sources. We estimate the mass and cooling ages of our WD and WD+MS candidates and compare those parameters with values from the literature. Section 4 discusses the identification of Extremely Low-Mass WD (ELM) candidates in our sample. We also discuss the completeness of our WD identification and its implications on WD space density estimates. A summary of our findings is encapsulated in Section 5.

\section{Data used in this study}

\begin{figure}
    \centering
    \includegraphics[width=\columnwidth,height=\textheight,keepaspectratio]{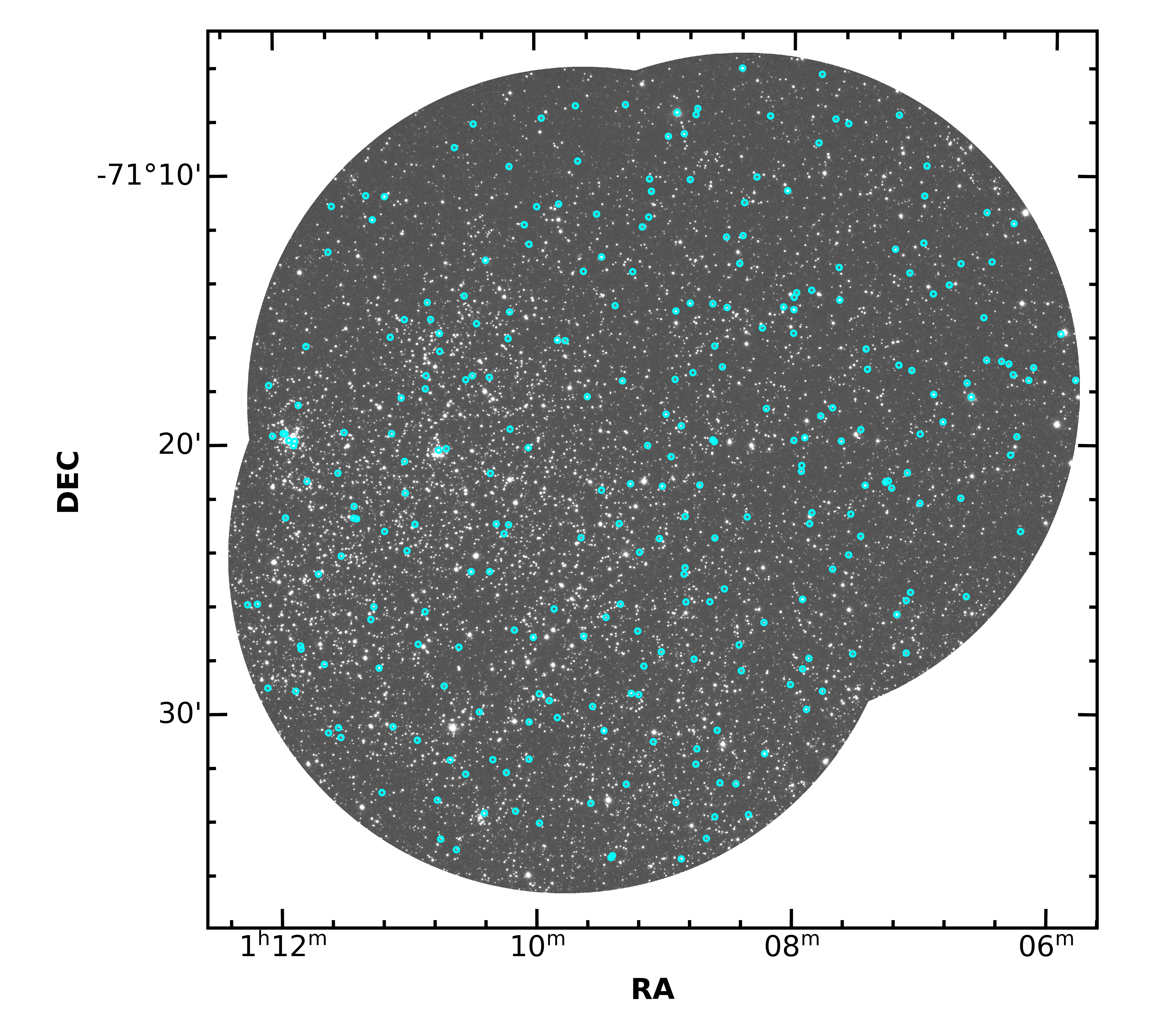}
    \caption{The greyscale mosaic of the three SMC fields observed by UVIT in the N245M filter. The image was convolved by a Gaussian kernel of 1.5\arcsec\ for better visualization. The sample of 273 sources in this study is given in cyan open circles.}
    \label{SMC}
\end{figure}

We made use of the UVIT point-source catalogue (hereafter, SMC-UVIT-1), which includes the observations taken at three fields with combined FoV of $\sim$ 40\arcmin\ towards the SMC (centred at $\alpha_{2000}=\ \sim17.285^{\circ}$ \& $\delta_{2000}=\ \sim-71.329^{\circ}$), between 2017 December 31 and 2018 January 1 \citep{devaraj2023uvit}. The three fields were shifted $\sim$ 6\arcmin\ in an orthogonal direction from each other, with some level of overlap between them. The first field was chosen far away from the centre of the SMC to avoid the bright central region \citep{tandon2020additional}. Figure \ref{SMC} shows the grey-scale mosaic of the three SMC fields observed by UVIT in the N245M filter. UVIT has an impressive $\sim$ 1.5\arcsec\ spatial resolution which is better than the $\sim$ 5\arcsec\ spatial resolution of GALEX or $\sim$ 3\arcsec\ spatial resolution of Ultraviolet Imaging Telescope (UIT) \citep{stecher1997ultraviolet}. For further details on the exposure time and the information on different filters used, refer to \citet{devaraj2023uvit}. SMC-UVIT-1 has a combined total of 11,241 objects detected across 7 UVIT filters viz \textit{F154W, F169M, F172M, N245M, N263M, N279N and N219M}, with a limiting magnitude of $\sim$21 mag.  

We cross-matched the SMC-UVIT-1 and Gaia DR3 catalogue\footnote{\url{https://gea.esac.esa.int/archive/}} to complement our UV data with astrometric and optical photometric information. 
The cross-matching was performed employing a search radius of 1.5\arcsec, aligning with the spatial resolution of the UVIT instrument. This resulted in a source list of 10,847 objects having both UVIT and Gaia measurements (hereafter SMC-UVIT-Gaia). The average offset in angular separation between the corresponding UVIT and Gaia sources is around 0.1\arcsec, with $\sim$ 90\% of the sources matching within 0.4\arcsec.

We applied the following astrometric and photometric quality criteria to the Gaia DR3 catalogue, adapted from \citet{rebassa2021white}:
\begin{itemize}
    \item $I_{BP}/\sigma_{I_{BP}} \geq 10$
    \item $I_{RP}/\sigma_{I_{RP}} \geq 10$
    \item $I_{G}/\sigma_{I_{G}} \geq 10$
    \item $\varpi/\sigma_{\varpi}\geq 3$
    \item $\varpi > 0$
\end{itemize}

where $\varpi$ is the parallax in arcseconds, $I_{G}$ , $I_{BP}$ and $I_{RP}$ are the fluxes in  the bandpass filters $G$, $G_{BP}$ and $G_{RP}$, respectively, and the $\sigma$ values are the standard errors of the corresponding parameters. Further, we applied the following photometric quality criteria to the available UVIT magnitudes:
\begin{itemize}
    \item $M_{X}/\sigma_{M_{X}} \geq 10$
\end{itemize}
where $M_{X}$ and $\sigma_{M_{X}}$ represent the magnitude and the error in magnitude in each of the seven UVIT filters. It is to be noted that not all stars have UVIT magnitudes in all 7 filters.

From the sample of 10,847 sources, 273 sources qualified the given criteria, with 87.9\% of the sources having Renormalised Unit Weight Error (RUWE) $\leq$ 1.4. About 76\% of 273 sources have data in at least 3 UV filters in addition to the 3 Gaia optical filters. This dataset comprising 273 sources having robust photometric and astrometric information is used in this study.

\section{Analysis and Results}

\subsection{Do the sources belong to SMC or the Milky Way?}

\begin{figure}
    \centering
    \includegraphics[width=\columnwidth,height=\textheight,keepaspectratio]{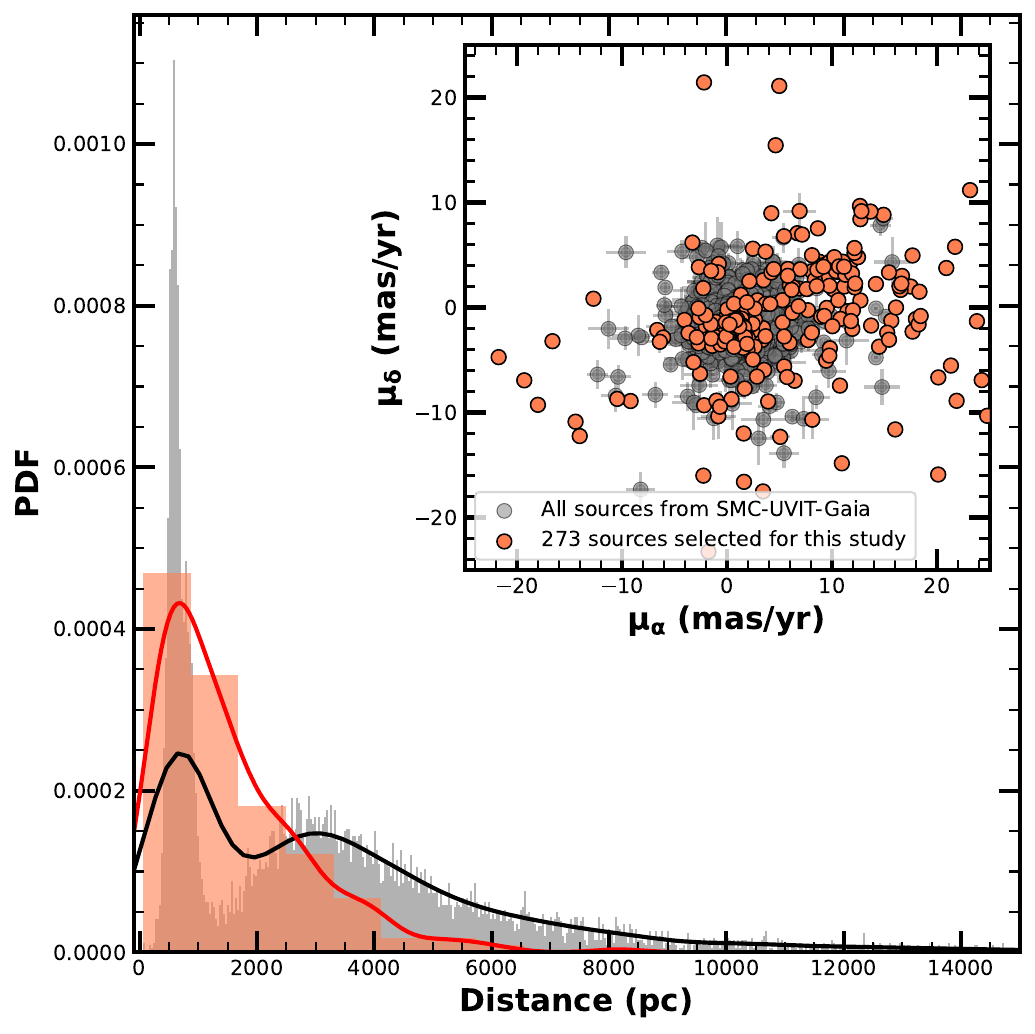}
    \caption{Gaia DR3 distance \citep{bailer2021estimating} distribution plot of the 273 sources used in this study. The 273 sources used in this study are given in orange and the 10,847 sources in the SMC-UVIT-Gaia in grey markers. Kernel density estimates (KDE) of the distributions are overlaid for a better visual representation. The red and black lines are the KDE plots of the 273 sources and SMC-UVIT-Gaia sources, respectively. In the inset: Scatter plot with error bars of proper motion in RA (mas/yr) vs proper motion in Dec (mas/yr). For most sources, error bars are of the size of the markers.}
    \label{DistHist} 
\end{figure}

A critical aspect of our analysis is to discern whether our sample of 273 sources belong to SMC or if they are Milky Way (MW) sources projected in the line of sight to the SMC. To accomplish this, we leveraged the distance data provided by Gaia DR3 for all the sources \citep{bailer2021estimating}, which is depicted as a histogram in Figure \ref{DistHist}. Our 273 sources lie within the distance range of 50 pc -- 6500 pc, with the maximum distance recorded at 8176 ± 769 pc. Notably, the SMC is situated at a distance of 61.9 ± 0.6 kpc \citep{de2015clustering}, establishing a stark contrast between our sources' distances and the known distance of SMC. In addition to the distance values, from the plot of proper motion in Dec ($\mu_{\delta}$)  vs proper motion in RA ($\mu_{\alpha}$) given in Figure \ref{DistHist} (inset), it is clear that the majority of 273 sources have higher values compared to the rest of the SMC-UVIT-Gaia sources. This confirms that the 273 sources belong to the MW, and is found to be in the projection of the SMC. 

\begin{figure}
    \centering
    \includegraphics[width=\columnwidth,height=\textheight,keepaspectratio]{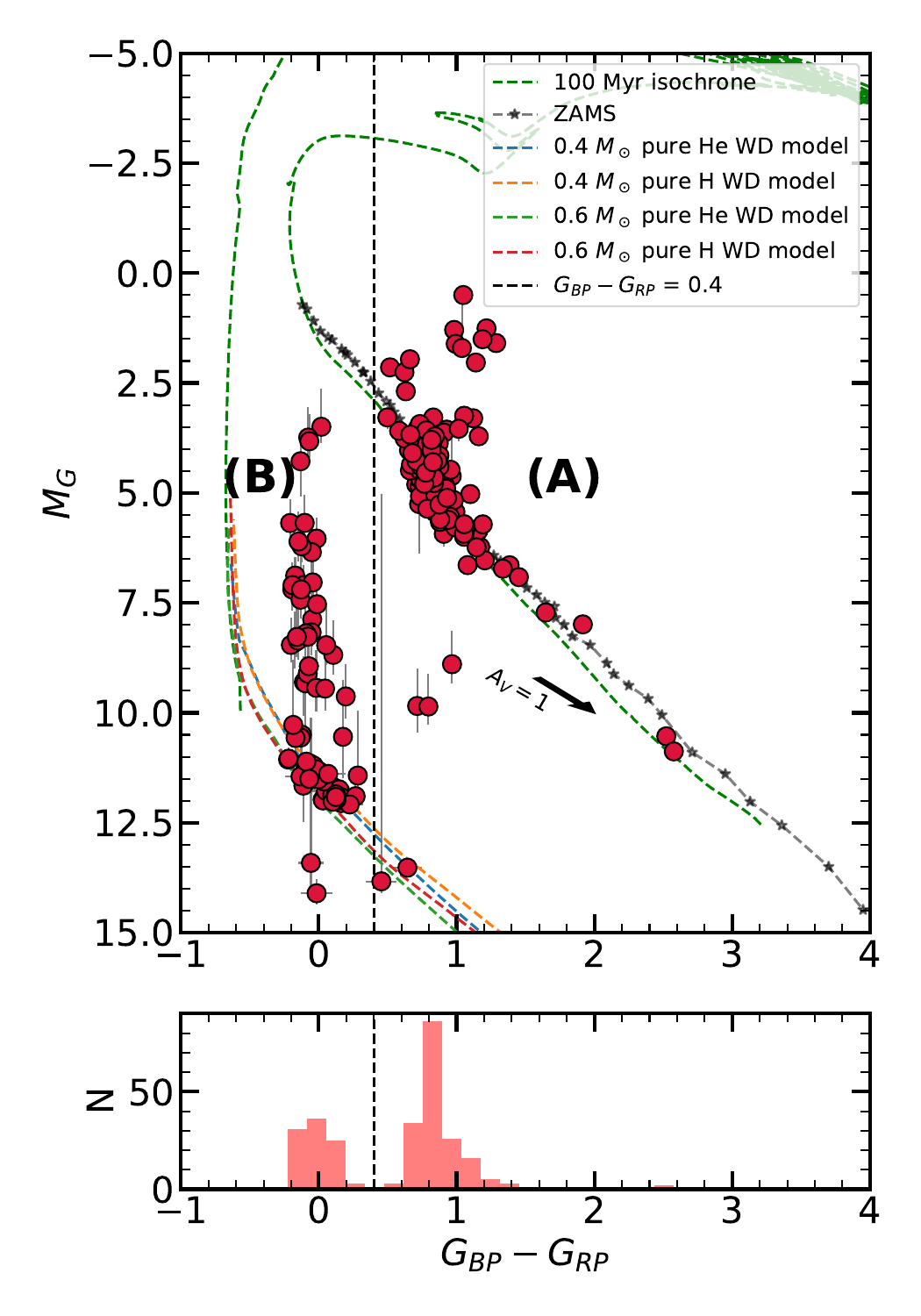}
    \caption{Top panel: Extinction corrected Gaia CMD of the 273 sources (red-filled circles) with $G_{BP}-G_{RP}$ on the X-axis and $M_{G}$ on the Y-axis. The black vertical line that divides the CMD into two populations is $G_{BP}-G_{RP} = 0.4$. The black arrow shows the reddening vector for Av = 1 mag. Bottom panel: The number distribution of the 273 sources in the $G_{BP}-G_{RP}$ colour. }
    \label{fig1} 
\end{figure}

Furthermore, to have an understanding of the types of sources in our sample, we performed a SIMBAD crossmatch of the 273 sources and found that 24 sources have previous classifications in the literature. We note that six sources are classified as WD candidates (WD*\_candidate) by \citet{gentile2021catalogue}. Out of these six WD candidates, Gaia DR2 4690619477160130944 was also listed as a WD candidate in \citet{jimenez2018white}. 13 are classified as `star'. Two sources are classified as `$RGB^*$\_candidate', one as RR Lyrae Variable (RRLyr), one as a high proper motion star ($PM^*$) and one as a Classical Cepheid Variable (deltaCep). We also crossmatched our list of 273 sources with the WD catalogue provided by \citet{jackim2024galex} and found 12 matches. One of the sources classified as a WD candidate by \citet{jackim2024galex} was also classified as a WD candidate by \citet{gentile2021catalogue}. The list of these objects is given in Table~\ref{tab:Simbad}. 

\begin{figure*}
    \centering
    \includegraphics[width=\textwidth,height=\textheight,keepaspectratio]{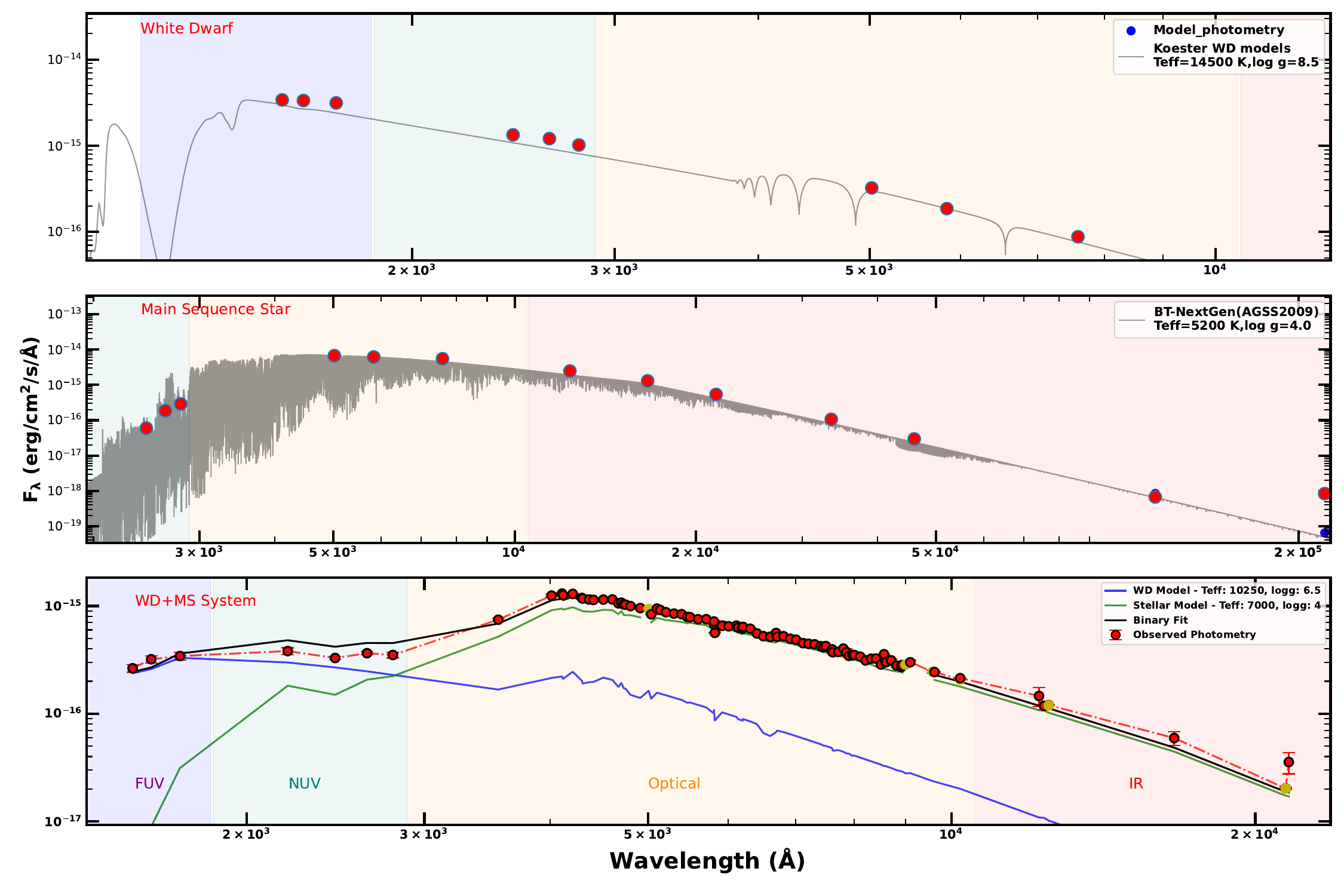}
    \caption{A representative sample of SEDs of WD (Gaia DR3 4690625700572457088), MS star (Gaia DR3 4690614052626726016), WD+MS (Gaia DR3 4690657483318685312) over-plotted by their respective best-fitted models. The scaling of the X-axis and Y-axis of each plot are different.}
    \label{SED}
\end{figure*}

\subsection{Disentangling MS Sources and WD Sources}
As seen with the literature crossmatch, we expect to find MS, RGB and WD sources in our sample. With the exceptional astrometry from Gaia, we create a Gaia Color-Magnitude Diagram (CMD) and segregate the sample of MS/RGB and WD sources based on their positions in the CMD. 

We plot the absolute $G$ band magnitude $M_{G}$ against $G_{BP}-G_{RP}$ color for 273 sources in Figure \ref{fig1} along with the Zero Age Main Sequence (ZAMS) \citep{pecaut2013intrinsic}, a 100 Myr isochrone from Modules for Experiments in Stellar
Astrophysics (MESA)\footnote{\url{https://waps.cfa.harvard.edu/MIST/}} \citep{choi2016mesa,dotter2016mesa} and WD cooling tracks from \citet{bedard2020spectral}. The CMD shows a diverse population including WDs, MS stars, and potential WD+MS star systems. A clear distinction can be seen in the CMD and we use $G_{BP}-G_{RP} > $ 0.4 mag as the cutoff to separate MS and WD sources (shown as regions A and B, respectively). The bottom panel in Figure \ref{fig1} shows a clear distinction between the population at $G_{BP}-G_{RP} = $ 0.4. This distinction aligns with the expected optical brightness contrast between MS stars and WDs, enabling a reliable differentiation. This approach is based on the fact that WDs will be brighter in bluer bands and MS stars in redder. WDs emit a significant amount of their radiation in the blue and ultraviolet parts of the spectrum. As a result, they tend to be brighter in bluer bands (in this case, BP mag) compared to MS stars. The cutoff yielded 95 potential WD sources in region B, and 178 potential MS stars in region A, exhibiting characteristics consistent with MS stars. The SMC is characterized by a notably low foreground extinction value of E(B-V) = 0.02 mag \citep{hutchings1982international}, representing the maximum potential extinction that our sources could encounter in this line of sight. Hence, for the purposes of our analysis, we are considering an extinction value of $A_{V}$~=~0.062 mag, which corresponds to E(B-V) = 0.02 mag for the sources.

\subsection{SED analysis of UV-Bright MS and WD sources}

To characterise the 178 potential MS sources, we employed a Python-based routine tailored to fit observed SEDs, similar to the SED routine used in \citet{arun2021clustering}, \citet{bhattacharyya2022decoding} and \citet{shridharan2022emission}. The SED for a target object is constructed using photometric data from various sources spanning wavelengths from far-UV to far-IR, including available photometry from \textit{UVIT, UCAC4, SDSS, Gaia, 2MASS, AKARI IRC, AKARI FIS, WISE, and IRAS.} SED templates used for the fitting process are derived from the BT-NextGen (AGSS2009) model grid \citep{2011ASPC..448...91A, 2012RSPTA.370.2765A}, considering the parameters of effective temperature ($800K \leq T_{eff} \leq 70000K$), surface gravity ($-0.5\leq $logg$ \leq 6$) and metallicity ($-4\leq [Fe/H] \leq 0.5$).

The fitting procedure involved a normalization process of observed and model photometry, utilizing user-specified bands. The determination of the best-fit model relied on minimizing the chi-squared parameter ($\chi^2$), computed from fitting the UVIT bands, Gaia \textit{G}, \textit{$G_{BP}$}, and \textit{$G_{RP}$} band photometry, thereby giving the most suitable model corresponding to the lowest $\chi^2$. Subsequently, stellar parameters such as $T_{eff}$ and $logg$ are estimated from the SED fitting. Upon visual inspection of fit quality, our analysis revealed that 161 sources out of 178, exhibited very good fit with theoretical models, indicative of their classification as UV-bright MS stars (provided in Table \ref{tab:MS}). Out of 161 sources, $\sim 89\%$ of the sources have $logg$ ranging from 4 to 5.5. The $T_{eff}$ of these sources range from 4500~K to 6500~K. Conversely, 16 stars encountered fitting discrepancies that can be primarily attributed to problematic photometric data or the presence of a companion and will be further analysed in Section \ref{wd+ms}.

\begin{table*}
\centering
\caption{List of the 43 single white dwarfs identified in the present study. The WDs marked in bold were previously detected in the literature as WD candidates.}
\label{tab:WD}
\resizebox{\textwidth}{!}{%
\begin{tabular}{lllllllllllllllllllll}
\hline
\multicolumn{1}{c}{\textbf{\begin{tabular}[c]{@{}c@{}}RA\\ (deg)\end{tabular}}} & \multicolumn{1}{c}{\textbf{\begin{tabular}[c]{@{}c@{}}DE\\ (deg)\end{tabular}}} & \multicolumn{1}{c}{\textbf{\begin{tabular}[c]{@{}c@{}}Distance \\ (pc)\end{tabular}}} & \multicolumn{2}{c}{\textbf{\begin{tabular}[c]{@{}c@{}}N263M\\ (mag)\end{tabular}}} & \multicolumn{2}{c}{\textbf{\begin{tabular}[c]{@{}c@{}}N245M\\ (mag)\end{tabular}}} & \multicolumn{2}{c}{\textbf{\begin{tabular}[c]{@{}c@{}}N279N\\ (mag)\end{tabular}}} & \multicolumn{2}{c}{\textbf{\begin{tabular}[c]{@{}c@{}}F172M\\ (mag)\end{tabular}}} & \multicolumn{2}{c}{\textbf{\begin{tabular}[c]{@{}c@{}}F169M\\ (mag)\end{tabular}}} & \multicolumn{2}{c}{\textbf{\begin{tabular}[c]{@{}c@{}}F154W\\ (mag)\end{tabular}}} & \multicolumn{2}{c}{\textbf{\begin{tabular}[c]{@{}c@{}}N219M\\ (mag)\end{tabular}}} & \textbf{\begin{tabular}[c]{@{}l@{}}Teff\\ (K)\end{tabular}} & \textbf{logg} & \textbf{\begin{tabular}[c]{@{}l@{}}Cooling Age\\ (Gyr)\end{tabular}} & \multicolumn{1}{c}{\textbf{\begin{tabular}[c]{@{}c@{}}WD Mass\\ ($M_{\odot}$)\end{tabular}}} \\ \cline{4-17}
\multicolumn{1}{c}{}                                                            & \multicolumn{1}{c}{}                                                            & \multicolumn{1}{c}{}                                                                  & \multicolumn{1}{c}{\textbf{AB}}        & \multicolumn{1}{c}{\textbf{Error}}        & \multicolumn{1}{c}{\textbf{AB}}        & \multicolumn{1}{c}{\textbf{Error}}        & \multicolumn{1}{c}{\textbf{AB}}        & \multicolumn{1}{c}{\textbf{Error}}        & \multicolumn{1}{c}{\textbf{AB}}        & \multicolumn{1}{c}{\textbf{Error}}        & \multicolumn{1}{c}{\textbf{AB}}        & \multicolumn{1}{c}{\textbf{Error}}        & \multicolumn{1}{c}{\textbf{AB}}        & \multicolumn{1}{c}{\textbf{Error}}        & \multicolumn{1}{c}{\textbf{AB}}        & \multicolumn{1}{c}{\textbf{Error}}        & \multicolumn{1}{c}{}                                        &               & \textbf{}                                                            & \textbf{}                                                                                    \\ \hline
17.69418                                                                        & -71.55454                                                                       & $542^{+47}_{-113}$                                                                    & 19.779                                 & 0.051                                     & 19.895                                 & 0.047                                     & 19.598                                 & 0.088                                     & 19.876                                 & 0.082                                     & 20.168                                 & 0.078                                     & 19.929                                 & 0.066                                     &                                        &                                           & 12500                                                       & 6.75          & 0.1                                                                  & 0.35                                                                                         \\
17.66857                                                                        & -71.52984                                                                       & $2581^{+803}_{-538}$                                                                  & 18.644                                 & 0.03                                      & 18.679                                 & 0.027                                     & 18.8                                   & 0.061                                     & 18.801                                 & 0.05                                      & 18.761                                 & 0.041                                     & 18.944                                 & 0.042                                     & 18.619                                 & 0.055                                     & 13250                                                       & 8.5           &                                                                      & ELM                                                                                          \\
\textbf{17.80608$\ssymbol{2}$}                                                  & \textbf{-71.47253}                                                              & \textbf{$1173^{+336}_{-219}$}                                                         & \textbf{18.959}                        & \textbf{0.035}                            & \textbf{18.944}                        & \textbf{0.03}                             & \textbf{19.067}                        & \textbf{0.069}                            & \textbf{19.078}                        & \textbf{0.057}                            & \textbf{19}                            & \textbf{0.046}                            & \textbf{19.099}                        & \textbf{0.045}                            & \textbf{18.957}                        & \textbf{0.064}                            & \textbf{13250}                                              & \textbf{8.25} & \textbf{}                                                            & \textbf{ELM}                                                                                 \\
17.22698                                                                        & -71.55645                                                                       & $1008^{+195}_{-148}$                                                                  & 19.01                                  & 0.036                                     & 19.171                                 & 0.034                                     & 19.056                                 & 0.069                                     & 19.158                                 & 0.059                                     & 19.399                                 & 0.055                                     & 19.575                                 & 0.056                                     &                                        &                                           & 11250                                                       & 6.75          &                                                                      & ELM                                                                                          \\
17.14111                                                                        & -71.54422                                                                       & $492^{+36}_{-41}$                                                                     &                                        &                                           & 21.548                                 & 0.101                                     &                                        &                                           & 21.334                                 & 0.161                                     &                                        &                                           & 21.786                                 & 0.155                                     &                                        &                                           & 11250                                                       & 6.75          & 0.54                                                                 & 0.51                                                                                         \\
17.38993                                                                        & -71.49697                                                                       & $514^{+178}_{-43}$                                                                    & 20.409                                 & 0.049                                     & 20.579                                 & 0.046                                     & 20.53                                  & 0.098                                     & 20.463                                 & 0.077                                     & 20.733                                 & 0.07                                      & 20.846                                 & 0.072                                     & 20.496                                 & 0.129                                     & 11000                                                       & 6.5           & 0.24                                                                 & 0.42                                                                                         \\
17.51403                                                                        & -71.50647                                                                       & $1427^{+1407}_{-344}$                                                                 & 19.687                                 & 0.036                                     & 19.724                                 & 0.031                                     & 19.652                                 & 0.064                                     & 19.745                                 & 0.077                                     & 19.787                                 & 0.066                                     & 20.065                                 & 0.07                                      & 19.783                                 & 0.093                                     & 11750                                                       & 6.75          &                                                                      & ELM                                                                                          \\
17.58956                                                                        & -71.41328                                                                       & $448^{+36}_{-28}$                                                                     & 20.018                                 & 0.033                                     & 20.129                                 & 0.031                                     & 20.227                                 & 0.069                                     & 20.03                                  & 0.062                                     & 20.435                                 & 0.064                                     & 20.279                                 & 0.055                                     & 20.202                                 & 0.084                                     & 12000                                                       & 6.5           & 0.29                                                                 & 0.43                                                                                         \\
17.28988                                                                        & -71.47199                                                                       & $435^{+16}_{-21}$                                                                     &                                        &                                           & 21.758                                 & 0.08                                      &                                        &                                           &                                        &                                           &                                        &                                           & 22.164                                 & 0.183                                     & 21.344                                 & 0.176                                     & 10250                                                       & 6.5           & 0.45                                                                 & 0.4                                                                                          \\
17.33562                                                                        & -71.43358                                                                       & $461^{+25}_{-26}$                                                                     & 20.463                                 & 0.042                                     & 20.472                                 & 0.036                                     & 20.523                                 & 0.104                                     & 20.732                                 & 0.082                                     & 20.698                                 & 0.056                                     & 20.982                                 & 0.066                                     & 20.366                                 & 0.082                                     & 12750                                                       & 8.5           & 0.35                                                                 & 0.47                                                                                         \\
17.05608                                                                        & -71.44488                                                                       & $450^{+17}_{-37}$                                                                     & 20.155                                 & 0.035                                     & 20.294                                 & 0.033                                     & 20.085                                 & 0.064                                     & 20.651                                 & 0.072                                     & 20.696                                 & 0.056                                     & 20.963                                 & 0.064                                     & 20.252                                 & 0.081                                     & 11250                                                       & 7.25          & 0.26                                                                 & 0.41                                                                                         \\
17.57628                                                                        & -71.3837                                                                        & $863^{+765}_{-477}$                                                                   & 19.375                                 & 0.025                                     & 19.443                                 & 0.022                                     & 19.307                                 & 0.044                                     & 19.688                                 & 0.043                                     & 19.743                                 & 0.035                                     & 19.896                                 & 0.038                                     & 19.349                                 & 0.052                                     & 11500                                                       & 7.5           &                                                                      & ELM                                                                                          \\
17.51368                                                                        & -71.33661                                                                       & $4663^{+1146}_{-1737}$                                                                & 18.087                                 & 0.013                                     & 18.177                                 & 0.013                                     & 18.182                                 & 0.026                                     & 18.336                                 & 0.023                                     & 18.488                                 & 0.02                                      & 18.54                                  & 0.02                                      & 18.175                                 & 0.025                                     & 12750                                                       & 8.25          &                                                                      &                                                                                              \\
16.7787                                                                         & -71.46324                                                                       & $398^{+66}_{-27}$                                                                     & 20.857                                 & 0.083                                     & 20.669                                 & 0.067                                     & 20.808                                 & 0.15                                      & 21.097                                 & 0.103                                     & 21.182                                 & 0.086                                     & 21.493                                 & 0.102                                     &                                        &                                           & 11000                                                       & 6.75          & 0.51                                                                 & 0.56                                                                                         \\
16.77949                                                                        & -71.4307                                                                        & $480^{+27}_{-22}$                                                                     & 20.854                                 & 0.083                                     & 20.949                                 & 0.076                                     & 20.59                                  & 0.136                                     & 21.036                                 & 0.137                                     & 21.286                                 & 0.093                                     & 21.496                                 & 0.099                                     &                                        &                                           & 11000                                                       & 6.75          & 0.34                                                                 & 0.55                                                                                         \\
16.56812                                                                        & -71.32855                                                                       & $502^{+36}_{-32}$                                                                     & 20.567                                 & 0.073                                     & 20.764                                 & 0.07                                      & 20.557                                 & 0.134                                     & 20.761                                 & 0.121                                     & 21.022                                 & 0.105                                     & 20.943                                 & 0.105                                     &                                        &                                           & 12000                                                       & 7.5           & 0.33                                                                 & 0.57                                                                                         \\
\textbf{16.71119$\ssymbol{1}$}                                                  & \textbf{-71.31982}                                                              & \textbf{$1554^{+436}_{-278}$}                                                         & \textbf{18.639}                        & \textbf{0.03}                             & \textbf{18.767}                        & \textbf{0.028}                            & \textbf{18.618}                        & \textbf{0.055}                            & \textbf{18.784}                        & \textbf{0.049}                            & \textbf{18.867}                        & \textbf{0.039}                            & \textbf{19.071}                        & \textbf{0.044}                            & \textbf{18.782}                        & \textbf{0.053}                            & \textbf{12000}                                              & \textbf{7.25} & \textbf{}                                                            & \textbf{ELM}                                                                                 \\
16.98139                                                                        & -71.43035                                                                       & $5410^{+1724}_{-1585}$                                                                & 18.026                                 & 0.013                                     & 18.08                                  & 0.012                                     & 18.007                                 & 0.024                                     & 18.217                                 & 0.022                                     & 18.235                                 & 0.018                                     & 18.313                                 & 0.018                                     & 18.058                                 & 0.028                                     & 12250                                                       & 7.75          &                                                                      &                                                                                              \\
16.86897                                                                        & -71.39106                                                                       & $498^{+78}_{-24}$                                                                     & 21.671                                 & 0.087                                     & 21.585                                 & 0.074                                     & 21.358                                 & 0.194                                     & 21.672                                 & 0.135                                     & 21.909                                 & 0.103                                     & 21.892                                 & 0.106                                     &                                        &                                           & 10750                                                       & 6.5           & 0.55                                                                 & 0.54                                                                                         \\
\textbf{17.15645$\ssymbol{1}$}                                                  & \textbf{-71.33187}                                                              & \textbf{$1860^{+526}_{-416}$}                                                         & \textbf{18.827}                        & \textbf{0.019}                            & \textbf{18.855}                        & \textbf{0.017}                            & \textbf{18.877}                        & \textbf{0.036}                            & \textbf{18.869}                        & \textbf{0.03}                             & \textbf{18.884}                        & \textbf{0.024}                            & \textbf{18.931}                        & \textbf{0.024}                            & \textbf{18.773}                        & \textbf{0.032}                            & \textbf{14250}                                              & \textbf{9}    & \textbf{}                                                            & \textbf{ELM}                                                                                 \\
17.15341                                                                        & -71.33289                                                                       & $557^{+66}_{-20}$                                                                     & 20.5                                   & 0.042                                     & 20.709                                 & 0.04                                      & 20.435                                 & 0.078                                     & 20.732                                 & 0.074                                     & 20.822                                 & 0.06                                      & 20.957                                 & 0.063                                     & 20.335                                 & 0.066                                     & 12500                                                       & 8.5           & 0.14                                                                 & 0.59                                                                                         \\
\textbf{17.31568$\ssymbol{1}$}                                                  & \textbf{-71.35908}                                                              & \textbf{$1889^{+512}_{-298}$}                                                         & \textbf{18.928}                        & \textbf{0.02}                             & \textbf{19.056}                        & \textbf{0.019}                            & \textbf{19.084}                        & \textbf{0.04}                             & \textbf{19.128}                        & \textbf{0.034}                            & \textbf{19.228}                        & \textbf{0.028}                            & \textbf{19.274}                        & \textbf{0.028}                            & \textbf{18.976}                        & \textbf{0.035}                            & \textbf{12500}                                              & \textbf{8}    & \textbf{}                                                            & \textbf{ELM}                                                                                 \\
\textbf{17.24697$\ssymbol{2}$}                                                  & \textbf{-71.31595}                                                              & \textbf{$1119^{+420}_{-238}$}                                                         & \textbf{18.691}                        & \textbf{0.018}                            & \textbf{18.8}                          & \textbf{0.016}                            & \textbf{18.748}                        & \textbf{0.034}                            & \textbf{18.814}                        & \textbf{0.029}                            & \textbf{18.888}                        & \textbf{0.024}                            & \textbf{18.918}                        & \textbf{0.024}                            & \textbf{18.763}                        & \textbf{0.032}                            & \textbf{13500}                                              & \textbf{8.5}  & \textbf{}                                                            & \textbf{ELM}                                                                                 \\
16.90776                                                                        & -71.33224                                                                       & $2377^{+824}_{-555}$                                                                  & 18.897                                 & 0.02                                      & 18.995                                 & 0.018                                     & 18.939                                 & 0.038                                     & 19.099                                 & 0.033                                     & 19.127                                 & 0.027                                     & 19.238                                 & 0.028                                     & 19.051                                 & 0.046                                     & 11750                                                       & 6.5           &                                                                      & ELM                                                                                          \\
17.13819                                                                        & -71.2866                                                                        & $524^{+68}_{-25}$                                                                     & 20.821                                 & 0.05                                      & 20.806                                 & 0.042                                     & 20.857                                 & 0.112                                     & 20.911                                 & 0.076                                     & 21.068                                 & 0.078                                     & 21.227                                 & 0.072                                     & 20.666                                 & 0.128                                     & 11500                                                       & 7.25          & 0.31                                                                 & 0.57                                                                                         \\
17.12949                                                                        & -71.24981                                                                       & $1534^{+575}_{-443}$                                                                  & 19.256                                 & 0.023                                     & 19.293                                 & 0.021                                     & 19.249                                 & 0.044                                     & 19.463                                 & 0.039                                     & 19.492                                 & 0.031                                     & 19.601                                 & 0.033                                     & 19.332                                 & 0.049                                     & 12250                                                       & 7.75          &                                                                      & ELM                                                                                          \\
17.45362                                                                        & -71.18575                                                                       & $470^{+30}_{-205}$                                                                    & 20.415                                 & 0.049                                     & 20.543                                 & 0.045                                     & 20.503                                 & 0.096                                     & 20.743                                 & 0.086                                     & 21                                     & 0.082                                     & 21.085                                 & 0.088                                     & 20.487                                 & 0.084                                     & 11750                                                       & 7.75          & 0.31                                                                 & 0.64                                                                                         \\
16.62831                                                                        & -71.28139                                                                       & $1379^{+657}_{-273}$                                                                  & 19.584                                 & 0.046                                     & 19.669                                 & 0.042                                     & 19.529                                 & 0.083                                     & 20.028                                 & 0.086                                     & 20.165                                 & 0.071                                     & 20.175                                 & 0.073                                     & 19.921                                 & 0.089                                     & 11250                                                       & 6.75          &                                                                      & ELM                                                                                          \\
16.48553                                                                        & -71.26473                                                                       & $2612^{+646}_{-459}$                                                                  & 17.784                                 & 0.02                                      & 17.834                                 & 0.018                                     & 17.84                                  & 0.038                                     & 17.675                                 & 0.029                                     & 17.748                                 & 0.023                                     & 17.823                                 & 0.025                                     &                                        &                                           & 15000                                                       & 9             &                                                                      &                                                                                              \\
17.09648                                                                        & -71.18481                                                                       & $189^{+27}_{-22}$                                                                     & 21.591                                 & 0.118                                     & 21.602                                 & 0.073                                     & 21.198                                 & 0.18                                      & 21.456                                 & 0.167                                     & 21.804                                 & 0.155                                     & 22.258                                 & 0.192                                     & 21.056                                 & 0.154                                     & 11750                                                       & 8.25          & 1.91                                                                 & 1.26                                                                                         \\
17.81519                                                                        & -71.43458                                                                       & $631^{+38}_{-152}$                                                                    & 20.001                                 & 0.041                                     & 20.018                                 & 0.036                                     & 19.745                                 & 0.066                                     & 20.129                                 & 0.065                                     & 20.095                                 & 0.053                                     & 20.159                                 & 0.052                                     & 19.824                                 & 0.095                                     & 13000                                                       & 8.5           & 0.09                                                                 & 0.29                                                                                         \\
17.84723                                                                        & -71.38002                                                                       & $1478^{+335}_{-365}$                                                                  & 19.663                                 & 0.034                                     & 19.803                                 & 0.032                                     & 19.497                                 & 0.059                                     & 20.064                                 & 0.068                                     & 19.913                                 & 0.047                                     & 20.073                                 & 0.05                                      & 19.552                                 & 0.057                                     & 14000                                                       & 9.5           &                                                                      & ELM                                                                                          \\
17.85196                                                                        & -71.37223                                                                       & $1075^{+430}_{-270}$                                                                  & 20.142                                 & 0.043                                     & 20.252                                 & 0.039                                     & 20.559                                 & 0.109                                     & 20.409                                 & 0.075                                     & 20.27                                  & 0.056                                     & 20.367                                 & 0.057                                     & 20.153                                 & 0.075                                     & 12500                                                       & 7.75          &                                                                      & ELM                                                                                          \\
18.00839                                                                        & -71.32826                                                                       & $337^{+30}_{-19}$                                                                     & 20.173                                 & 0.045                                     & 20.221                                 & 0.039                                     & 20.128                                 & 0.079                                     &                                        &                                           &                                        &                                           &                                        &                                           &                                        &                                           & 12000                                                       & 7.75          & 0.57                                                                 & 0.65                                                                                         \\
17.96764                                                                        & -71.33433                                                                       & $2231^{+485}_{-404}$                                                                  & 18.749                                 & 0.023                                     & 18.821                                 & 0.02                                      & 18.667                                 & 0.04                                      & 18.952                                 & 0.053                                     & 19.041                                 & 0.043                                     & 19.132                                 & 0.045                                     &                                        &                                           & 11000                                                       & 6.5           &                                                                      & ELM                                                                                          \\
\textbf{17.9879$\ssymbol{1}$}                                                   & \textbf{-71.32667}                                                              & \textbf{$1729^{+829}_{-335}$}                                                         & \textbf{19.066}                        & \textbf{0.026}                            & \textbf{19.249}                        & \textbf{0.025}                            & \textbf{19.092}                        & \textbf{0.051}                            & \textbf{}                              & \textbf{}                                 & \textbf{}                              & \textbf{}                                 & \textbf{}                              & \textbf{}                                 & \textbf{}                              & \textbf{}                                 & \textbf{11500}                                              & \textbf{8.5}  & \textbf{}                                                            & \textbf{ELM}                                                                                 \\
17.98428                                                                        & -71.32692                                                                       & $1788^{+381}_{-380}$                                                                  & 19.307                                 & 0.03                                      &                                        &                                           & 19.156                                 & 0.07                                      &                                        &                                           &                                        &                                           &                                        &                                           &                                        &                                           & 11000                                                       & 6.5           &                                                                      & ELM                                                                                          \\
17.97703                                                                        & -71.33103                                                                       & $874^{+234}_{-312}$                                                                   & 19.334                                 & 0.042                                     & 19.426                                 & 0.027                                     & 19.245                                 & 0.073                                     & 19.284                                 & 0.062                                     & 19.391                                 & 0.051                                     & 19.366                                 & 0.051                                     &                                        &                                           & 14250                                                       & 6.5           &                                                                      & ELM                                                                                          \\
17.95815                                                                        & -71.30935                                                                       & $882^{+424}_{-149}$                                                                   & 19.415                                 & 0.03                                      & 19.617                                 & 0.029                                     & 19.651                                 & 0.066                                     & 19.778                                 & 0.077                                     & 19.93                                  & 0.065                                     & 19.98                                  & 0.067                                     &                                        &                                           & 11500                                                       & 7             &                                                                      & ELM                                                                                          \\
17.75881                                                                        & -71.3053                                                                        & $3081^{+918}_{-667}$                                                                  & 18.608                                 & 0.021                                     & 18.591                                 & 0.019                                     & 18.679                                 & 0.041                                     & 18.773                                 & 0.035                                     & 18.713                                 & 0.027                                     & 18.761                                 & 0.027                                     & 18.503                                 & 0.035                                     & 13500                                                       & 8.75          &                                                                      &                                                                                              \\
17.63595                                                                        & -71.24246                                                                       & $500^{+37}_{-65}$                                                                     & 20.761                                 & 0.047                                     & 20.902                                 & 0.045                                     & 20.611                                 & 0.082                                     & 21.067                                 & 0.084                                     & 21.363                                 & 0.078                                     & 21.465                                 & 0.088                                     & 20.613                                 & 0.125                                     & 12250                                                       & 8.75          & 0.39                                                                 & 0.46                                                                                         \\
17.54867                                                                        & -71.16241                                                                       & $422^{+35}_{-23}$                                                                     & 21.12                                  & 0.068                                     & 21.095                                 & 0.058                                     & 20.9                                   & 0.115                                     & 21.382                                 & 0.162                                     & 21.328                                 & 0.124                                     & 21.599                                 & 0.142                                     &                                        &                                           & 10750                                                       & 6.5           & 0.52                                                                 & 0.44                                                                                         \\
17.40706$\ssymbol{3}$                                                           & -71.45349                                                                       & $269^{+16}_{-26}$                                                                     & 18.76                                  & 0.019                                     & 18.91                                  & 0.017                                     & 18.757                                 & 0.034                                     & 19.112                                 & 0.033                                     & 19.132                                 & 0.027                                     & 19.245                                 & 0.028                                     & 18.887                                 & 0.043                                     & 11500                                                       & 7.25          & 0.24                                                                 & 0.49                                                                                         \\ \hline
\end{tabular}%
}

\begin{tablenotes}
\item$\ssymbol{1}$ \citet{gentile2021catalogue}, $\ssymbol{2}$ \citet{jackim2024galex},
 $\ssymbol{3}$ WD showing IR excess
           
\end{tablenotes}
\end{table*}

\begin{table*}
\centering
\caption{List of the 13 WD+MS candidates identified from the present study with the parameters of both the MS and WD components}
\label{tab:WDMS}
\resizebox{\textwidth}{!}{%
\begin{tabular}{lllllllllllllllllllllll}
\hline
\multicolumn{1}{c}{\textbf{\begin{tabular}[c]{@{}c@{}}RA\\ (deg)\end{tabular}}} & \multicolumn{1}{c}{\textbf{\begin{tabular}[c]{@{}c@{}}DE\\ (deg)\end{tabular}}} & \multicolumn{1}{c}{\textbf{\begin{tabular}[c]{@{}c@{}}Distance \\ (pc)\end{tabular}}} & \multicolumn{2}{c}{\textbf{\begin{tabular}[c]{@{}c@{}}N263M\\ (mag)\end{tabular}}} & \multicolumn{2}{c}{\textbf{\begin{tabular}[c]{@{}c@{}}N245M\\ (mag)\end{tabular}}} & \multicolumn{2}{c}{\textbf{\begin{tabular}[c]{@{}c@{}}N279N\\ (mag)\end{tabular}}} & \multicolumn{2}{c}{\textbf{\begin{tabular}[c]{@{}c@{}}F172M\\ (mag)\end{tabular}}} & \multicolumn{2}{c}{\textbf{\begin{tabular}[c]{@{}c@{}}F169M\\ (mag)\end{tabular}}} & \multicolumn{2}{c}{\textbf{\begin{tabular}[c]{@{}c@{}}F154W\\ (mag)\end{tabular}}} & \multicolumn{2}{c}{\textbf{\begin{tabular}[c]{@{}c@{}}N219M\\ (mag)\end{tabular}}} & \textbf{\begin{tabular}[c]{@{}l@{}}MS\_Teff\\ (K)\end{tabular}} & \textbf{MS\_logg} & \textbf{\begin{tabular}[c]{@{}l@{}}WD\_Teff\\ (K)\end{tabular}} & \textbf{WD\_logg} & \textbf{\begin{tabular}[c]{@{}l@{}}WD Cooling Age\\ (Gyr)\end{tabular}} & \multicolumn{1}{c}{\textbf{\begin{tabular}[c]{@{}c@{}}WD Mass\\ ($M_{\odot}$)\end{tabular}}} \\ \cline{1-17}
\multicolumn{1}{c}{}                                                            & \multicolumn{1}{c}{}                                                            & \multicolumn{1}{c}{}                                                                  & \multicolumn{1}{c}{\textbf{AB}}        & \multicolumn{1}{c}{\textbf{Error}}        & \multicolumn{1}{c}{\textbf{AB}}        & \multicolumn{1}{c}{\textbf{Error}}        & \multicolumn{1}{c}{\textbf{AB}}        & \multicolumn{1}{c}{\textbf{Error}}        & \multicolumn{1}{c}{\textbf{AB}}        & \multicolumn{1}{c}{\textbf{Error}}        & \multicolumn{1}{c}{\textbf{AB}}        & \multicolumn{1}{c}{\textbf{Error}}        & \multicolumn{1}{c}{\textbf{AB}}        & \multicolumn{1}{c}{\textbf{Error}}        & \multicolumn{1}{c}{\textbf{AB}}        & \multicolumn{1}{c}{\textbf{Error}}        &                                                                 &                   & \multicolumn{1}{c}{}                                            &                   & \textbf{}                                                               & \textbf{}                                                                                    \\ \hline
17.58841                                                                        & -71.29290                                                                       & $430^{+27}_{-28}$                                                                     & 21.227                                 & 0.059                                     & 21.372                                 & 0.054                                     &                                        &                                           & 21.581                                 & 0.129                                     & 21.919                                 & 0.117                                     & 21.904                                 & 0.117                                     & 20.976                                 & 0.105                                     & 7000                                                            & 4.5               & 13750                                                           & 9.5               & 0.50                                                                    & 0.46                                                                                         \\
17.27154                                                                        & -71.51881                                                                       & $413^{+21}_{-20}$                                                                     & 21.18                                  & 0.098                                     & 21.216                                 & 0.087                                     &                                        &                                           & 21.521                                 & 0.175                                     & 21.269                                 & 0.13                                      & 21.293                                 & 0.123                                     &                                        &                                           & 6800                                                            & 4                 & 12250                                                           & 6.5               & 0.50                                                                    & 0.47                                                                                         \\
17.00401                                                                        & -71.48322                                                                       & $428^{+26}_{-30}$                                                                     & 21.41                                  & 0.107                                     & 21.472                                 & 0.097                                     &                                        &                                           &                                        &                                           & 21.689                                 & 0.143                                     & 21.98                                  & 0.169                                     &                                        &                                           & 7000                                                            & 4                 & 11000                                                           & 6.5               & 0.44                                                                    & 0.43                                                                                         \\
17.29824                                                                        & -71.40148                                                                       & $464^{+19}_{-12}$                                                                     & 21.782                                 & 0.129                                     & 21.789                                 & 0.112                                     & 21.485                                 & 0.204                                     & 21.988                                 & 0.145                                     & 21.95                                  & 0.162                                     & 21.949                                 & 0.119                                     &                                        &                                           & 5800                                                            & 4                 & 11250                                                           & 6.5               & 0.52                                                                    & 0.49                                                                                         \\
17.67229                                                                        & -71.33692                                                                       & $397^{+18}_{-29}$                                                                     & 21.394                                 & 0.081                                     & 21.664                                 & 0.076                                     & 21.109                                 & 0.134                                     & 21.433                                 & 0.166                                     & 21.823                                 & 0.156                                     &                                        &                                           &                                        &                                           & 6700                                                            & 4                 & 11750                                                           & 6.5               & 0.50                                                                    & 0.43                                                                                         \\
17.33770                                                                        & -71.38375                                                                       & $1022^{+349}_{-235}$                                                                  &                                        &                                           & 21.482                                 & 0.097                                     &                                        &                                           & 22.074                                 & 0.173                                     &                                        &                                           &                                        &                                           &                                        &                                           & 7000                                                            & 4                 & 10250                                                           & 6.5               &                                                                         & ELM                                                                                          \\
16.96834                                                                        & -71.38353                                                                       & $452^{+26}_{-33}$                                                                     & 21.476                                 & 0.112                                     & 21.456                                 & 0.058                                     & 21.129                                 & 0.173                                     & 21.956                                 & 0.139                                     & 22.014                                 & 0.17                                      & 22.131                                 & 0.137                                     &                                        &                                           & 7000                                                            & 4.5               & 12000                                                           & 7.75              & 0.55                                                                    & 0.49                                                                                         \\
17.51080                                                                        & -71.21059                                                                       & $429^{+18}_{-12}$                                                                     &                                        &                                           & 21.961                                 & 0.121                                     &                                        &                                           &                                        &                                           & 22.926                                 & 0.22                                      &                                        &                                           &                                        &                                           & 7000                                                            & 4                 & 10500                                                           & 6.5               & 0.59                                                                    & 0.42                                                                                         \\
17.38071                                                                        & -71.19201                                                                       & $327^{+46}_{-31}$                                                                     & 21.135                                 & 0.071                                     & 21.32                                  & 0.069                                     & 21.213                                 & 0.133                                     & 21.822                                 & 0.149                                     & 22.325                                 & 0.196                                     & 22.123                                 & 0.135                                     & 21.043                                 & 0.15                                      & 7000                                                            & 4.5               & 13250                                                           & 9.5               & 0.49                                                                    & 0.32                                                                                         \\
17.95762                                                                        & -71.46028                                                                       & $429^{+26}_{-26}$                                                                     & 21.352                                 & 0.106                                     & 21.298                                 & 0.09                                      &                                        &                                           & 21.579                                 & 0.18                                      & 21.779                                 & 0.165                                     &                                        &                                           &                                        &                                           & 7000                                                            & 4                 & 14500                                                           & 9.5               & 0.57                                                                    & 0.55                                                                                         \\
17.79306                                                                        & -71.38791                                                                       & $464^{+10}_{-42}$                                                                     & 21.603                                 & 0.119                                     & 21.507                                 & 0.099                                     &                                        &                                           & 21.925                                 & 0.208                                     &                                        &                                           &                                        &                                           &                                        &                                           & 7000                                                            & 4                 & 11000                                                           & 6.5               & 0.47                                                                    & 0.47                                                                                         \\
17.71227                                                                        & -71.29982                                                                       & $422^{+41}_{-22}$                                                                     & 21.118                                 & 0.057                                     & 21.295                                 & 0.054                                     & 21.195                                 & 0.178                                     & 21.333                                 & 0.161                                     & 21.385                                 & 0.094                                     & 21.473                                 & 0.095                                     &                                        &                                           & 6900                                                            & 5                 & 12500                                                           & 7.75              & 0.55                                                                    & 0.58                                                                                         \\
17.77915                                                                        & -71.26768                                                                       & $431^{+33}_{-14}$                                                                     &                                        &                                           & 21.762                                 & 0.111                                     &                                        &                                           &                                        &                                           & 22.699                                 & 0.233                                     &                                        &                                           &                                        &                                           & 7000                                                            & 5                 & 10500                                                           & 6.5               & 0.55                                                                    & 0.52                                                                                         \\ \hline
\end{tabular}%
}
\end{table*}

To characterize the WD sources, we fit the Koester WD models for pure H atmospheres \citep{koester2010white} to the 95 sources lying in region B of the Gaia CMD. The fitting process involves a systematic comparison between observed SEDs and theoretical WD models. By assessing various parameters, such as $T_{eff}$ and $logg$, the routine determines the most suitable WD model that best matches the observed data. Subsequently, the fitted SEDs underwent a visual classification, categorizing them based on the goodness of their fit. 43 of 95 sources showed a very good SED fit, indicating them being highly probable WD candidates. Checking against the existing literature, 37 out of these 43 WDs are found to be new detections. Four sources (Gaia DR3 4690615216552295808, Gaia DR3 4690615560150914560, Gaia DR3 4690613017528995840, Gaia DR3 4690659300093357184) were previously classified as highly probable WD candidates by \citet{gentile2021catalogue} and two sources (Gaia DR3 4690616659661394304, Gaia DR3 4690551758417486336) were classified as WD candidates in \citet{jackim2024galex}. These are marked by boldface in Table \ref{tab:WD}. Furthermore, the stellar parameters encompassing critical factors such as effective temperature and surface gravity for the identified 43 WDs are tabulated in Table \ref{tab:WD}. The SED analysis resulted in 52 sources which couldn't be fitted with a single WD model. Some of these sources showed optical or IR excess. These objects will be looked into in detail in the next section as potential WD+MS sources.

\subsection{Identifying WD+MS Binaries}
\label{wd+ms}

WD+MS binaries pose a unique challenge since the flux of one of the stars dominates the SED, making their identification in the HR diagram elusive. These systems will have similar magnitudes and colours as single stars. However, systems where both the WD and MS companion contribute significantly to the optical flux, form a bridge between the WD and MS loci in the CMD \citep{rebassa2021white}. Our initial analysis revealed that our sample may also include WD+MS binary systems since some objects lie between the WD locus denoted by the WD cooling tracks and MS loci indicated by the ZAMS in  Figure \ref{fig1} displaying the CMD.

Our strategy involved visually classifying SEDs showing optical and IR excess during WD model fitting or the sources that were not fitting well with the single WD model as potential binary systems. This resulted in the identification of 52 such sources. Utilizing the VO SED analyzing tool (VOSA\footnote{\url{http://svo2.cab.inta-csic.es/theory/vosa/}}; \citealp{bayo2008vosa}), we fitted binary models for these candidates. For the MS star component in our binary modelling, we utilized the BT-Settl-CIFIST models \citep{baraffe2015new}, offering a wide database of spectra across $T_{eff}$ ranging from 1,200 K to 7,000 K, with $logg$ within $4 \leq $logg$ \leq 5$. For H-rich WDs, we employed Koester (2010) WD evolution models \citep{koester2010white}, with $T_{eff}$ between 5,000 K and 80,000 K, and $logg$ within 6.5 $\leq$ $logg$ $\leq$ 9.5.

\begin{figure*}
    \centering
    \includegraphics[width=\textwidth,height=\textheight,keepaspectratio]{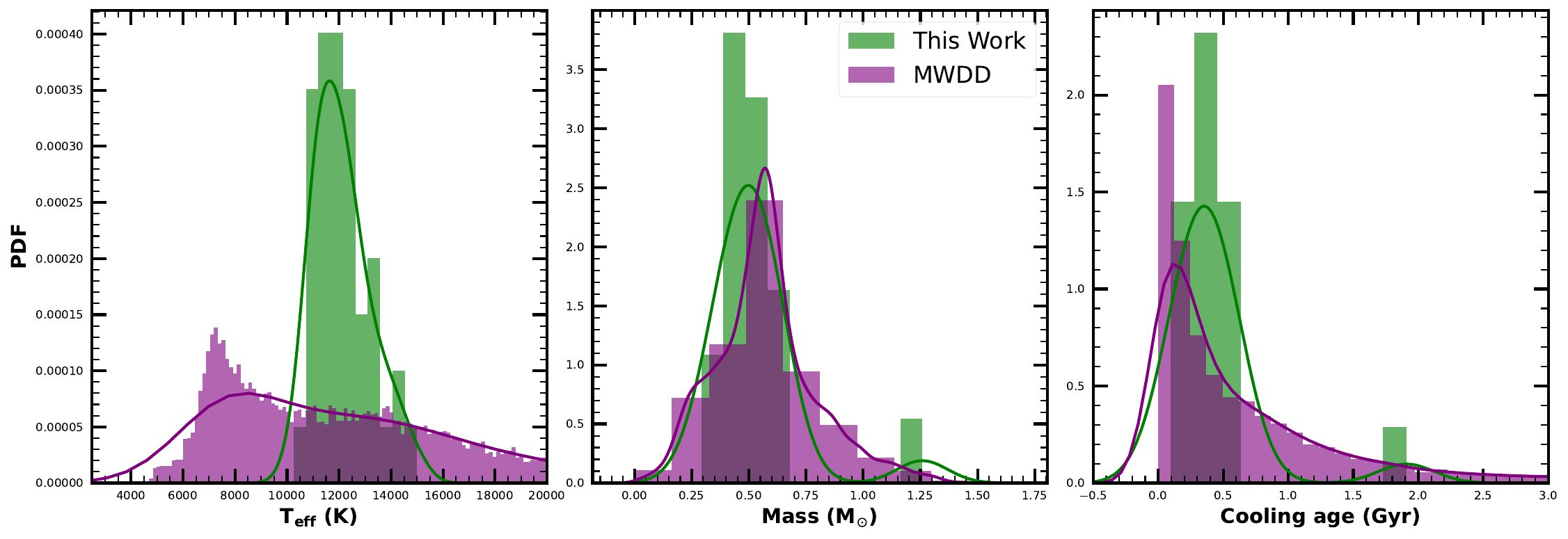}
    \caption{Comparison of $T_{eff}$ (K), mass (\(M_\odot\)) and the cooling age (Gyr) for single WDs identified in this work (green) and MWDD (purple). KDEs of the distributions are also overlaid for a better visual representation.}
    \label{hist_single_WD}
\end{figure*}

Among these 52 sources, visual inspection revealed 13 stars exhibiting good fit, implying them being potential binary systems. A representative sample of a well-fitted WD+MS system is shown in the bottom panel of Figure \ref{SED}. Table \ref{tab:WDMS} presents the parameters of these selected WD+MS binary candidates. Notably, we refrained from relying on reduced chi-square parameters provided by VOSA. Existing literature \citep{rebassa2021white,nayak2022hunting} has cautioned against their use due to potentially misleading outcomes. Instead, we used the prevalent and accepted parameter called visual goodness-of-fit ($Vgf_{b}$) provided by VOSA along with visually inspecting the goodness of fit in our analysis. All 13 candidates have $Vgf_{b}\ <15$, which is usually considered as a proxy for well-fitted SEDs. 

\subsection{Estimation of Mass and the cooling age of WDs and WD+MS candidates}
\label{mass}

The determination of WD masses is crucial in understanding their physical properties and evolution. While the mass can be derived during SED fitting, its accuracy heavily depends on the derived $logg$ of the WD. However, since SED is essentially a low-resolution spectrum, the $logg$ derived from SED fitting may not be precise, leading to unreliable mass estimates.

To overcome this limitation, we adopt a method commonly used in the literature (e.g., \citealp{nayak2022hunting,karinkuzhi2024mass}), which utilizes the model cooling curves of WDs to estimate their masses. For this purpose, we utilize the `\texttt{WD\_models}' Python open-source package \footnote{\url{https://github.com/SihaoCheng/WD_models}}, which facilitates the conversion between WD optical photometry and physical parameters \citep{cheng2020double}. Specifically, we employ the cooling models developed by \citet{bedard2020spectral}\footnote{Available at \url{http://www.astro.umontreal.ca/~bergeron/CoolingModels/}}, which account for WDs with thick H atmospheres and CO cores. `\texttt{WD\_models}' estimates the mass and other parameters by finding the closest atmosphere grid and cooling model in the Gaia CMD.

By applying this methodology, we try to estimate the masses of 43 single WDs using their Gaia photometry. Among these WDs, 19 falls within the mass range covered by the available models, which spans from 0.2~$M_{\odot}$ to 1.3~$M_{\odot}$. Consequently, we estimate the masses for these 18 WDs. However, the remaining 24 WDs have masses below 0.2~$M_{\odot}$ and exhibit luminosities exceeding the model evolutionary tracks. As a result, we cannot estimate the masses for these sources through this method. Furthermore, we estimate the mass of the WD component of the 13 WD+MS candidates identified in this paper using the same method. We find that 12 of 13 sources lie within the model range and we estimate the masses for them. One WD+MS candidate has a mass below 0.2~$M_{\odot}$ and its mass cannot be calculated for the same reason mentioned earlier. Masses of the WD components range between 0.3~$M_{\odot}$ to 0.6~$M_{\odot}$.

In addition to the WD masses, we can also estimate the cooling age of these WDs from the model tracks using the `\texttt{WD\_models}' package. The cooling age of the single WDs ranges from 0.1 Gyr to 1.9 Gyr whereas for the WD components of the WD+MS binaries, cooling ages range from 0.2 Gyr to 0.6 Gyr. The mass and cooling age of single WDs and the WD+MS components are given in Table \ref{tab:WD} and Table \ref{tab:WDMS}, respectively.

\subsection{Comparison with existing WD and WD+MS catalogues}

It is important to compare the derived values of our sample to the values present in the literature on WDs. We made use of the Montreal White Dwarf Database\footnote{\url{https://www.montrealwhitedwarfdatabase.org/home.html}} \citep{dufour2016montreal}, a collection of confirmed white dwarfs in the literature. We selected single DA white dwarfs with at least one spectra available as our bonafide sample of WDs. We also removed WDs which show IR excess or have a debris disk around them. There were 18448 WDs which qualified for this criterion (hereafter, MWDD). 

Figure \ref{hist_single_WD} shows the comparison of $T_{eff}$, mass and the cooling age (Gyr) for single WDs in this study (green) with those found in the MWDD (purple). It is evident from the distributions that we identify hotter and slightly less massive WDs than MWDD. Considering that the majority of MWDD WDs are identified using optical surveys such as SDSS, this study using UVIT enhanced the sensitivity to the hotter WDs. We found the cooling ages of our WDs to be similar to the WDs identified in the literature. 

\begin{figure}
    \centering
    \includegraphics[width=\columnwidth,height=\textheight,keepaspectratio]{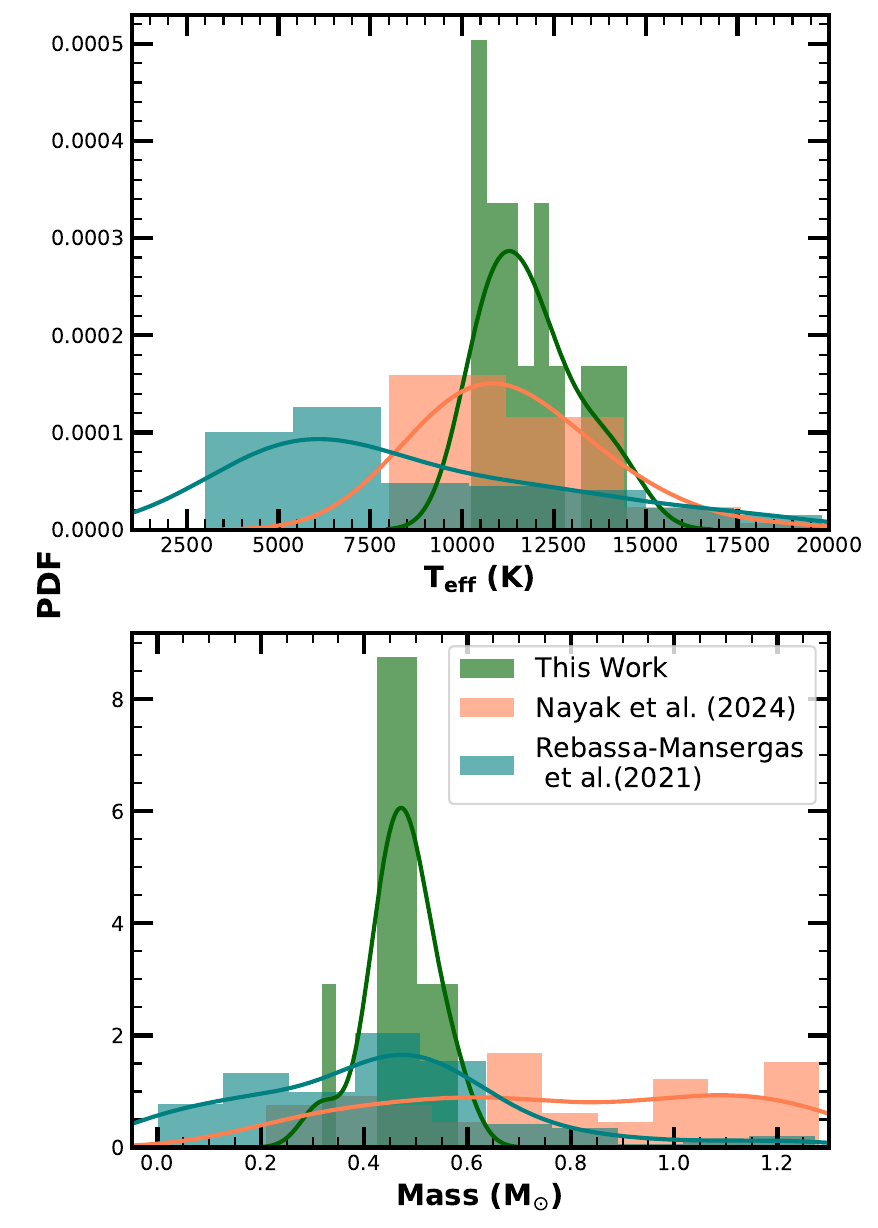}
    \caption{Comparison of $T_{eff}$ (K) and mass (\(M_\odot\)) of the WD component of the WD+MS candidates identified in this work with the works of \citet{rebassa2021white} and \citet{nayak2022hunting}.}
    \label{hist_WDMS}
\end{figure}

Furthermore, we compared the WD components of our WD+MS candidates with the most recent WD+MS catalogue provided by \citet{nayak2022hunting} and \citet{rebassa2021white}. Both these studies report WD+MS candidates within 100~pc. While \citet{rebassa2021white} uses the Gaia DR3 data for the identification of 112 WD+MS candidates, \citet{nayak2022hunting} uses UV data from GALEX GR6/7 combined with Gaia DR3 to identify  WD+MS candidates. \citet{rebassa2021white} targets sources that show up on Gaia optical CMD between the MS and the WD cooling sequence. Their 112 WD+MS candidates all belong to the region between MS and WDs on Gaia CMD, making them more sensitive to sources where the WD and the MS share similar contributions to the overall optical flux. On the other hand, \citet{nayak2022hunting} finds WD+MS candidates which reside in the WD locus in UV CMD and the MS locus in the optical CMD. In our work, we find our WD+MS binaries fall in the WD region of the Gaia CMD. Hence, the WD+MS population identified in this work is complementary to the sample detected by \citet{rebassa2021white} and \citet{nayak2022hunting}. Figure~\ref{hist_WDMS} shows the comparison between the $T_{eff}$ and mass of WDs in WD+MS systems identified in all three works. We find WDs with masses comparable to \citet{rebassa2021white} but less massive than the WDs found by \citet{nayak2022hunting}. We are identifying hotter WDs than both the previous studies. 
\section{Discussion}

In this section, we discuss the population of WDs that do not fall under the WD locus and present the potential identification of a sample of ELMs in our sample. Furthermore, with the identified sample of WDs, we calculate the WD space density and compare it with the previous estimates.

\subsection{Extremely-Low Mass White Dwarf Candidates}

\begin{figure}
    \centering
    \includegraphics[width=\columnwidth,keepaspectratio]{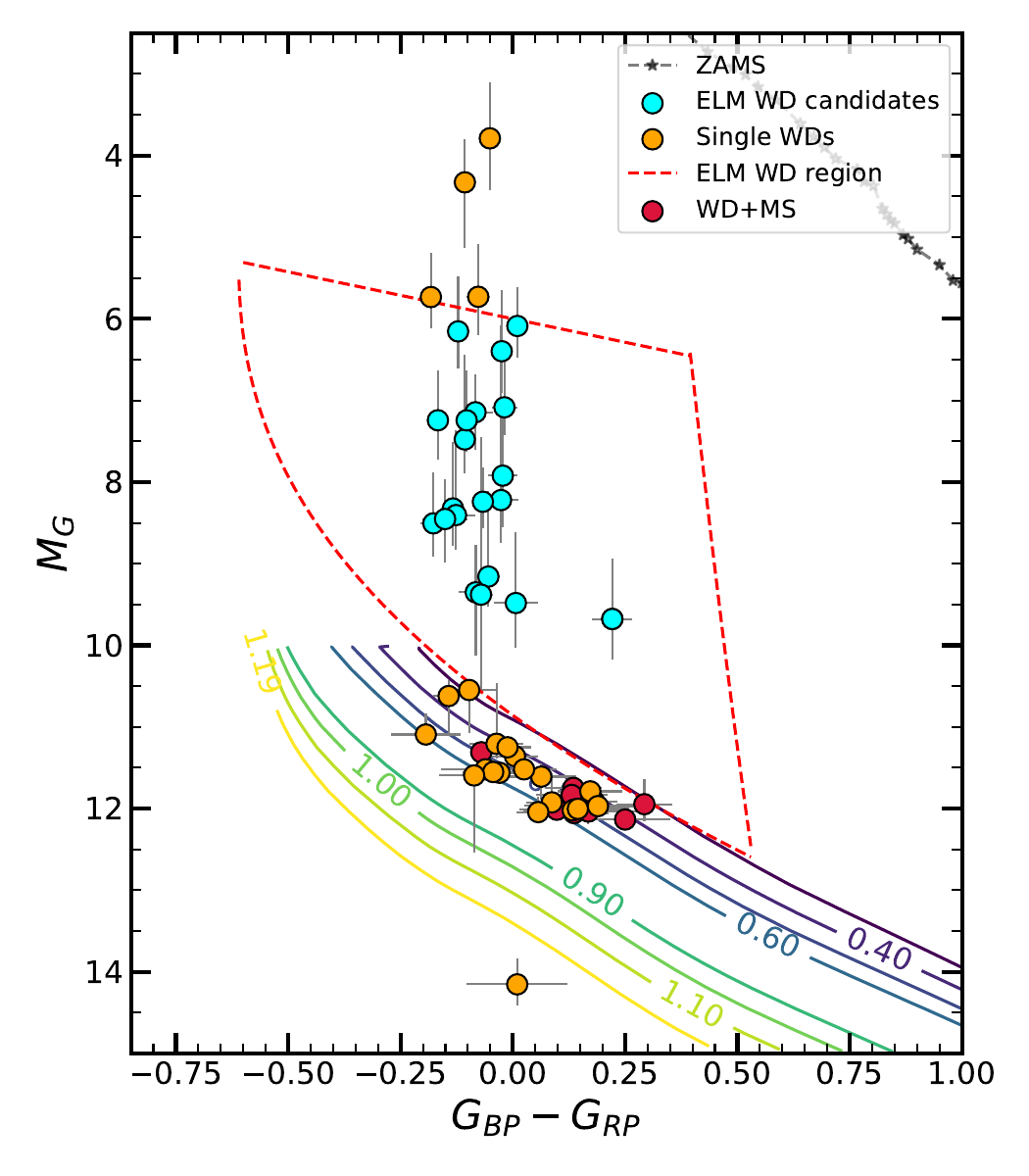}
    \caption{Extinction corrected Gaia CMD of single WDs and WD+MS candidates from our classification. The ELM region is marked with red dotted lines. ELM WD candidates are shown with cyan-filled circles. The remaining single WDs and WD+MS candidates are shown with orange and red-filled circles, respectively. The WD cooling models for different masses from \citet{bedard2020spectral} are also overlayed. The numbers on the cooling track denote their corresponding masses in \(M_\odot\). }
    \label{ELM}
\end{figure}

ELMs represent an exotic subset of WDs with masses below 0.3 \(M_\odot\), challenging the conventional understanding of single stellar evolution. These objects are expected to form via binary mass transfer from the WD progenitor during its red giant branch phase either through Roche lobe overflow or common envelope evolution \citep{istrate2014formation,istrate2014timescale,nandez2015recombination,li2019formation,pelisoli2019gaia,brown2022elm,nayak2022hunting,khurana2023dynamically}.  \citet{kilic2007future} established a minimal mass threshold of $\sim$ 0.3 \(M_\odot\) for WDs formed through standard evolutionary channels, considering the age of the Universe and associated timescales. Objects falling below this threshold necessitate alternative formation mechanisms, often linked to binary or multiple stellar systems. The mass and radius of a WD are inversely related \citep{parsons2017testing,bedard2017measurements}. This implies that the ELMs exhibit larger radii, making them brighter than their canonical counterparts. Consequently, they occupy an intermediate region in the CMD, positioned between the MS locus and the locus of Single WDs.

\citet{pelisoli2019gaia} defined a specific region in the Gaia CMD to identify the locus where ELMs are expected to appear. This region is defined by the equations:
\begin{equation}
M_{G} < 5.25 + 6.94\ (G_{BP}-G_{RP} + 0.61)^{1/2.32}
\end{equation}
\begin{equation}
M_{G} > 1.15\ (G_{BP}-G_{RP})\ +6.00
\end{equation}
\begin{equation}
M_{G} > -42.2\ (G_{BP}-G_{RP} )^{2}+83.8(G_{BP}-G_{RP} )-20.1
\end{equation}
where $M_{G}$ is the absolute G magnitude calculated using Gaia G band magnitude. 

As mentioned in Section~\ref{mass}, masses of 25 WDs (24 Single WDs and 1 WD+MS candidate) couldn't be found due to their higher luminosity. In Figure \ref{ELM}, we plotted the above-defined ELM region on the Gaia CMD. 20 of these 25 high luminosity WDs fall within the defined ELM region. These 20 WDs serve as strong candidates for ELMs and are marked by `ELM' in the `WD Mass' column of Tables \ref{tab:WD} and \ref{tab:WDMS}. They need to be followed up by spectroscopic observations to verify their candidacy. Follow-up studies of these 20 candidates will also allow us to constrain their physical parameters. The median distance of canonical mass WD sources from this study is $\sim$500 pc, whereas, for ELM candidates, the median distance is $\sim$1500 pc. We observe a higher percentage of ELMs in this study because ELMs are brighter than canonical mass WDs, resulting in better Gaia quality compared to WDs at a similar distance. Except for the extreme mass loss observed in high-metallicity stars \citep{kilic2007future}, ELM WDs are anticipated to originate from binary evolution, involving one or more instances of common envelope evolution \citep{li2019formation}. Therefore, investigating ELM systems can provide valuable insights into the wider realm of binary interactions.  

\subsection{Completeness of WD detection}

UV missions such as GALEX and UV-Excess survey (UVEX; \citealp{groot2009uv}) have been instrumental in studying the hot DA white dwarf population within 1kpc of the Sun. Based on the photometrically identified DA population with $T_{eff}$ > 10,000 K, the estimate of WD stellar density is $\sim$2.9-3.8 $\times$ 10$^{-4}$ pc$^{-3}$ within 1kpc \citep{verbeek2013determination}. Hence, considering a cone of radius 40' (FoV) at a distance of 1kpc, we can expect 48 DA WDs hotter than 10,000K. Using UVIT, we find 22 WDs hotter than 10,000K within 1kpc, providing a crude completeness estimate of 46\% percentage. 

However, it should also be noted that we are only considering the stars with good quality Gaia astrometry and photometry. As seen in Figure \ref{DistHist}, there is a large population of sources with distances less than 2kpc that did not satisfy our quality criteria. This population might have many new WDs with low-quality Gaia astrometry and photometry. Upon relaxing the $\varpi/\sigma_{\varpi}$ cutoff from $\geq$ 3 to $\geq$ 1, we additionally find 552 WD candidates showcasing an exceptional fit to WD models in the SED analysis. Distance estimates provided by Gaia DR3 would not be accurate for these distant low-luminosity sources. Based on the SED fit, we see that these 552 sources have T$_{eff}$ in the range of 7250K to 30,000K. Considering the UVIT sensitivity and degeneracies in mass, radius and $logg$, we estimate that they can be detected only to a distance limit of 8-10 kpc based on flux values of \citet{koester2010white} model of a 30,000K WD. Assuming 8~kpc as the maximum distance of 552 WD candidates, we can estimate an upper limit to the WD stellar density as $\sim$ 5-6 $\times$ 10$^{-4}$ pc$^{-3}$. Taking an alternative approach, if we consider the distances provided by Gaia DR3 for the low-quality WDs, then 181 WDs are present within 1kpc. Under the assumption that the WDs density is homogeneous and isotropic throughout 1kpc, the space density reaches $\sim$ 1.3 $\times$ 10$^{-3}$ pc$^{-3}$, higher than most of the literature estimates given in \citet[Table 3]{verbeek2013determination}. Our upper limit of global (up to 1 kpc) WD space density agrees with the local ($<$20pc) WD space density ($\sim$ 4 $\times$ 10$^{-3}$ pc$^{-3}$) estimated using Gaia EDR3 by \citet{gentile2021catalogue}. This showcases the need for deep UV photometry to identify Galactic WDs. However, it is well known throughout the literature that the WD population model changes with thin disc, thick disc and the halo \citep{bianchi2011catalogues}. Hence, it is clear that UV photometry, in combination with Gaia future releases, can be used to better estimate the space density of WDs, given the structure of the Galaxy. Large-scale WD detection studies combining UV and Gaia are necessary to accurately estimate the WD density in the Galaxy. Future Gaia data releases will constrain the astrometry of these low-quality WDs better. 

\section{Summary}

In this study, we have demonstrated the capabilities of UVIT to identify and characterise the WDs in the Galaxy. We made use of the UVIT point source catalogue published by \citet{devaraj2023uvit} observed towards the outskirts of the SMC region. Further, we crossmatched the UVIT photometry with Gaia DR3 and found that the 273 UV sources with good quality Gaia astrometry and photometry values belong to the Milky Way galaxy rather than to the SMC region. Using colour cuts and SED analysis, we identified 43 single WDs (given in Table \ref{tab:WD}) in a region in which only 6 WDs were previously detected as WD candidates based on the Gaia DR3 analysis \citep{gentile2021catalogue,jackim2024galex}. In addition to 43 WDs, we report the identification of 13 WD+MS candidates (Table \ref{tab:WDMS}) and 161 UV bright MS stars (Table \ref{tab:MS}). 

To estimate the physical characteristics of the identified WDs in our sample, we follow a common method adopted in literature (\citealp{nayak2022hunting, karinkuzhi2024mass}) using the `\texttt{WD\_models}' open-source python package. The package estimates the mass and age of the WDs by finding the closest atmosphere grid and cooling model based on the position in the Gaia CMD. We find that the masses of WDs identified in this study range from 0.2~$M_{\odot}$ to 1.3~$M_{\odot}$ and the $T_{eff}$ from 10000~K to 15000~K. Cooling ages of these WDs range from 0.1~Gyr to 2~Gyr. In comparison to the bonafide sample of WDs (MWDD), we find that our sample of WDs has higher temperatures owing to the detection through UVIT. We note that the estimated mass and the cooling age of WDs identified in this work match well with the literature values from MWDD. Furthermore, we report the detection of 20 ELM candidates based on their position in the Gaia CMD and SED analysis. Further detailed studies and spectroscopic confirmation of these sources will lead to a better understanding of their formation processes. Given the limitations of Gaia DR3 distance measurements of optically faint WDs, we discuss the completeness and estimates of the WD space density through our sample. Back-of-the-envelope calculations show that the estimated WD space density within 1kpc can be two orders of magnitude larger than previous estimates.

Our results not only highlight the instrumental effectiveness of UVIT but also anticipate the potential of upcoming dedicated UV missions like INSIST. These missions are premier instruments for the systematic discovery of WDs and other diverse stellar systems, providing invaluable insights into the complex dynamics of binary systems within our galaxy. As we conclude this study within the current FoV, our analytical approach will be extended to unveil additional WD systems in other fields observed by UVIT \citep{2024asi..confO..46P,2023ApJS..264...40M,2020ApJS..247...47L}. The subsequent phase involves the critical follow-up step of spectroscopic surveys. This effective blend of follow-up observations and thorough analysis establishes UVIT and upcoming missions as crucial contributors to enhancing our understanding of stellar systems in the ultraviolet domain.

\begin{acknowledgements}

We thank the reviewer for their valuable comments and suggestions, which have improved the manuscript. We are grateful to the Centre for Research, CHRIST (Deemed to be University), Bangalore, for the research grant extended to carry out the current project through the SEED money project (SMSS-2335, 11/2023). We thank the SIMBAD database and the online VizieR library service for helping us with the literature survey and obtaining relevant data. This publication uses the data from the UVIT, which is part of the AstroSat mission of the ISRO, archived at the Indian Space Science Data Centre (ISSDC). This work
has made use of data from the European Space Agency (ESA) mission Gaia (\url{https://www.cosmos.esa.int/Gaia}), processed by the Gaia Data Processing and
Analysis Consortium (DPAC, \url{https://www.cosmos.esa.int/web/Gaia/dpac/consortium}). Funding for the DPAC has been provided by national institutions, in particular, the institutions participating in the Gaia Multilateral Agreement. This publication makes use of VOSA, developed under the Spanish Virtual Observatory (\url{https://svo.cab.inta-csic.es}) project funded by MCIN/AEI/10.13039/501100011033/ through grant PID2020-112949GB-I00. VOSA has been partially updated by using funding from the European Union's Horizon 2020 Research and Innovation Programme, under Grant Agreement $n^{\circ}$ 776403 (EXOPLANETS-A). 
\end{acknowledgements}

\bibliographystyle{aa}
\bibliography{example}

\begin{appendix}

\section{Crossmatch of 273 sources with SIMBAD}
\begin{table*}
\centering
\caption{List of 35 sources from the 273 sources which had a literature match.}
\label{tab:Simbad}
\resizebox{\textwidth}{!}{%
\begin{tabular}{@{}lllll@{}}
\toprule
\textbf{RA (deg)} & \textbf{DEC (deg)} & \textbf{ID}                    & \textbf{Object Type} & \textbf{Reference}                           \\ \midrule
17.324315622255   & -71.5451385177136  & Flo 562                        & Star                 &  \citet{brown2021gaia}           \\
17.7106502826721  & -71.2916943448284  & SSTISAGEMA J011050.63-711730.0 & Star                 &  \citet{boyer2011surveying}     \\
16.6796158981582  & -71.2218013983896  & OGLE SMC-CEP-4797              & deltaCep             &  \citet{soszynski2016ogle}       \\
17.1564487311332  & -71.3318689685798  & Gaia DR3 4690615216552295808   & WD*\_Candidate       &  \citet{gentile2021catalogue}    \\
16.7713529712995  & -71.4256416321197  & Flo 520                        & Star                 &  \citet{florsch1972etude}        \\
17.5136828428392  & -71.3366073814231  & Gaia DR2 4690569105783446016   & Star                 &  \citet{pelisoli2019gaia}        \\
17.3156836844279  & -71.3590784755199  & Gaia DR3 4690615560150914560   & WD*\_Candidate       &  \citet{gentile2021catalogue}    \\
17.0009874928717  & -71.2656175718428  & SSTISAGEMA J010800.18-711556.2 & RGB*\_Candidate      &  \citet{boyer2011surveying}      \\
17.4063434253079  & -71.2274672874512  & SSTISAGEMA J010937.59-711338.8 & RGB*\_Candidate      &  \citet{boyer2011surveying}      \\
17.0007632587125  & -71.250989567725   & Gaia DR2 4690619477160130944   & WD*\_Candidate       &  \citet{jimenez2018white,gentile2021catalogue}        \\
17.3720747932814  & -71.3629982597683  & SSTISAGEMA J010929.22-712146.6 & Star                 &  \citet{boyer2011surveying}      \\
17.9879044542788  & -71.326669251763   & Gaia DR3 4690659300093357184   & WD*\_Candidate       &  \citet{gentile2021catalogue}    \\
17.2080457654542  & -71.4322737820098  & SSTISAGEMA J010849.92-712556.5 & Star                 &  \citet{boyer2011surveying}      \\
17.7072269440675  & -71.2462944249463  & Gaia DR3 4690710225517000064   & WD*\_Candidate       &  \citet{gentile2021catalogue,jackim2024galex}    \\
17.200465789012   & -71.2471671891088  & V* AI Tuc                      & RRLyr                &  \citet{soszynski2016ogle}       \\
16.7111946726546  & -71.3198202304278  & Gaia DR3 4690613017528995840   & WD*\_Candidate       &  \citet{gentile2021catalogue}    \\
17.5947497588541  & -71.2204588186018  & Flo 580                        & Star                 &  \citet{cruzalebes2019catalogue} \\
16.6573235791628  & -71.3043469573808  & HD 6782                        & PM*                  &  \citet{brown2016gaia}           \\
17.2261501076935  & -71.1290625783677  & CD-71 52                       & Star                 &  \citet{cruzalebes2019catalogue} \\
16.7975558874761  & -71.4394252973841  & AzV 107F                       & Star                 &  \citet{zacharias2013fourth}     \\
16.9785321749041  & -71.3302517486784  & TYC 9139-2292-1                & Star                 &  \citet{steinmetz2020sixth}      \\
16.5374945763772  & -71.2857297762281  & [SSP2002] E        & Star                 &  \citet{sharpee2002bv}           \\
16.6198321911708  & -71.2204758069251  & SSTISAGEMA J010628.75-711313.9 & Star                 &  \citet{boyer2011surveying}      \\
16.5784454988714  & -71.1965697551032  & Flo 511                        & Star                 &  \citet{steinmetz2020sixth}      \\
17.3143667205009  & -71.4889395826173  & Gaia DR3 4690562719174050048   & WD*\_Candidate       & \citet{jackim2024galex}                                      \\
17.399210952441   & -71.305061697809   & Gaia DR3 4690616011131760384   & WD*\_Candidate       & \citet{jackim2024galex}                                      \\
17.1521486208823  & -71.3926117140705  & Gaia DR3 4690613949547518208   & WD*\_Candidate       & \citet{jackim2024galex}                                      \\
17.1569075977901  & -71.2475319062813  & Gaia DR3 4690618759910796544   & WD*\_Candidate       & \citet{jackim2024galex}                                      \\
17.1815225128217  & -71.3598163557577  & Gaia DR3 4690614327504623360   & WD*\_Candidate       & \citet{jackim2024galex}                                      \\
17.2538839850791  & -71.3606027456645  & Gaia DR3 4690614155705930112   & WD*\_Candidate       & \citet{jackim2024galex}                                      \\
17.8060768911678  & -71.4725313365736  & Gaia DR3 4690551758417486336   & WD*\_Candidate       & \citet{jackim2024galex}                                      \\
16.8059214822535  & -71.2131825819021  & Gaia DR3 4690631335569472512   & WD*\_Candidate       & \citet{jackim2024galex}                                      \\
17.2789677473114  & -71.1703183236006  & Gaia DR3 4690621508690057216   & WD*\_Candidate       & \citet{jackim2024galex}                                      \\
17.2469723278347  & -71.315952637468   & Gaia DR3 4690616659661394304   & WD*\_Candidate       & \citet{jackim2024galex}                                      \\
17.3713081454728  & -71.2185421124811  & Gaia DR3 4690620203019782400   & WD*\_Candidate       & \citet{jackim2024galex}                                      \\ \bottomrule
\end{tabular}%
}
\end{table*}
 
\section{The list of MS sources identified in this study}

\begin{landscape}
\begin{table}[h]
\centering
\caption{ List of the 161 MS stars identified from the present study}
\label{tab:MS}
\resizebox{\textwidth}{!}{%
\begin{tabular}{@{}llllllllllllllllllll@{}}
\toprule
\multicolumn{1}{c}{\textbf{Gaia DR3 ID}} & \multicolumn{1}{c}{\textbf{\begin{tabular}[c]{@{}c@{}}RA\\ (deg)\end{tabular}}} & \multicolumn{1}{c}{\textbf{\begin{tabular}[c]{@{}c@{}}DE\\ (deg)\end{tabular}}} & \multicolumn{1}{c}{\textbf{\begin{tabular}[c]{@{}c@{}}Distance \\ (pc)\end{tabular}}} & \multicolumn{2}{c}{\textbf{\begin{tabular}[c]{@{}c@{}}N263M\\ (mag)\end{tabular}}} & \multicolumn{2}{c}{\textbf{\begin{tabular}[c]{@{}c@{}}N245M\\ (mag)\end{tabular}}} & \multicolumn{2}{c}{\textbf{\begin{tabular}[c]{@{}c@{}}N279N\\ (mag)\end{tabular}}} & \multicolumn{2}{c}{\textbf{\begin{tabular}[c]{@{}c@{}}F172M\\ (mag)\end{tabular}}} & \multicolumn{2}{c}{\textbf{\begin{tabular}[c]{@{}c@{}}F169M\\ (mag)\end{tabular}}} & \multicolumn{2}{c}{\textbf{\begin{tabular}[c]{@{}c@{}}F154W\\ (mag)\end{tabular}}} & \multicolumn{2}{c}{\textbf{\begin{tabular}[c]{@{}c@{}}N219M\\ (mag)\end{tabular}}} & \multicolumn{1}{c}{\textbf{\begin{tabular}[c]{@{}c@{}}$T_{eff}$\\ (K)\end{tabular}}} & \multicolumn{1}{c}{\textbf{$logg$}} \\ \cmidrule(lr){5-18}
\multicolumn{1}{c}{}                     & \multicolumn{1}{c}{}                                                            & \multicolumn{1}{c}{}                                                            & \multicolumn{1}{c}{}                                                                  & \multicolumn{1}{c}{\textbf{AB}}        & \multicolumn{1}{c}{\textbf{Error}}        & \multicolumn{1}{c}{\textbf{AB}}        & \multicolumn{1}{c}{\textbf{Error}}        & \multicolumn{1}{c}{\textbf{AB}}        & \multicolumn{1}{c}{\textbf{Error}}        & \multicolumn{1}{c}{\textbf{AB}}        & \multicolumn{1}{c}{\textbf{Error}}        & \multicolumn{1}{c}{\textbf{AB}}        & \multicolumn{1}{c}{\textbf{Error}}        & \multicolumn{1}{c}{\textbf{AB}}        & \multicolumn{1}{c}{\textbf{Error}}        & \multicolumn{1}{c}{\textbf{AB}}        & \multicolumn{1}{c}{\textbf{Error}}        & \multicolumn{1}{c}{}                                                            & \multicolumn{1}{c}{}              \\ \midrule
4.69062518518192E+018                    & 16.57654                                                                        & -71.29023                                                                       & 507.751678                                                                            & 20.297                                 & 0.064                                     & 21.432                                 & 0.095                                     & 19.7                                   & 0.09                                      &                                        &                                           &                                        &                                           &                                        &                                           &                                        &                                           & 5400                                                                            & 5                                 \\
4.69056151658326E+018                    & 17.32432                                                                        & -71.54514                                                                       & 801.134888                                                                            & 19.262                                 & 0.04                                      & 20.908                                 & 0.075                                     & 18.391                                 & 0.05                                      &                                        &                                           &                                        &                                           &                                        &                                           &                                        &                                           & 4600                                                                            & 4.5                               \\
4.69066191142658E+018                    & 17.86988                                                                        & -71.32645                                                                       & 2866.5105                                                                             & 20.41                                  & 0.049                                     & 21.052                                 & 0.058                                     & 20.18                                  & 0.081                                     &                                        &                                           &                                        &                                           &                                        &                                           &                                        &                                           & 6500                                                                            & 3.5                               \\
4.6907112219494E+018                     & 17.59475                                                                        & -71.22046                                                                       & 643.002869                                                                            & 14.677                                 & 0.003                                     & 15.562                                 & 0.004                                     & 14.54                                  & 0.005                                     & 18.268                                 & 0.023                                     & 19.282                                 & 0.028                                     & 19.61                                  & 0.033                                     & 15.712                                 & 0.013                                     & 6600                                                                            & 5                                 \\
4.69062542570005E+018                    & 16.73223                                                                        & -71.24065                                                                       & 874.526184                                                                            & 21.832                                 & 0.13                                      &                                        &                                           & 21.208                                 & 0.181                                     &                                        &                                           &                                        &                                           &                                        &                                           &                                        &                                           & 5400                                                                            & 4                                 \\
4.69056725465948E+018                    & 17.36359                                                                        & -71.44183                                                                       & 1165.23694                                                                            & 19.1                                   & 0.021                                     & 20.186                                 & 0.031                                     & 19.185                                 & 0.042                                     &                                        &                                           &                                        &                                           &                                        &                                           & 20.677                                 & 0.097                                     & 6100                                                                            & 4                                 \\
4.69061721372258E+018                    & 17.44166                                                                        & -71.27038                                                                       & 1316.94849                                                                            & 19.371                                 & 0.025                                     & 20.555                                 & 0.038                                     & 18.96                                  & 0.037                                     &                                        &                                           &                                        &                                           &                                        &                                           & 21.058                                 & 0.101                                     & 5900                                                                            & 4.5                               \\
4.69066205316384E+018                    & 17.77782                                                                        & -71.32738                                                                       & 2212.08789                                                                            & 19.087                                 & 0.027                                     & 19.847                                 & 0.033                                     & 19.058                                 & 0.049                                     &                                        &                                           &                                        &                                           &                                        &                                           & 19.908                                 & 0.067                                     & 6800                                                                            & 4                                 \\
4.69056811365289E+018                    & 17.58723                                                                        & -71.35238                                                                       & 6141.80762                                                                            & 21.393                                 & 0.079                                     & 21.762                                 & 0.111                                     & 21                                     & 0.163                                     &                                        &                                           &                                        &                                           &                                        &                                           &                                        &                                           & 7600                                                                            & 5                                 \\
4.69054941767336E+018                    & 17.35431                                                                        & -71.5906                                                                        & 1954.15002                                                                            & 20.919                                 & 0.087                                     &                                        &                                           &                                        &                                           &                                        &                                           &                                        &                                           &                                        &                                           &                                        &                                           & 6200                                                                            & 5                                 \\
4.69062178356771E+018                    & 17.48629                                                                        & -71.13251                                                                       & 2494.0332                                                                             & 20.166                                 & 0.061                                     & 21.452                                 & 0.096                                     & 19.742                                 & 0.091                                     &                                        &                                           &                                        &                                           &                                        &                                           &                                        &                                           & 6100                                                                            & 5                                 \\
4.69056694111955E+018                    & 17.21182                                                                        & -71.41501                                                                       & 277.851501                                                                            & 21.561                                 & 0.117                                     & 21.623                                 & 0.075                                     & 21.22                                  & 0.181                                     & 22.647                                 & 0.288                                     &                                        &                                           &                                        &                                           &                                        &                                           & 10800                                                                           & 0                                 \\
4.69056621956637E+018                    & 17.09981                                                                        & -71.4748                                                                        & 4935.03955                                                                            & 20.619                                 & 0.046                                     & 20.992                                 & 0.046                                     & 20.419                                 & 0.075                                     &                                        &                                           &                                        &                                           &                                        &                                           &                                        &                                           & 7400                                                                            & 5                                 \\
4.69056079502881E+018                    & 17.21656                                                                        & -71.5914                                                                        & 579.67865                                                                             &                                        &                                           &                                        &                                           &                                        &                                           &                                        &                                           & 22.023                                 & 0.184                                     &                                        &                                           &                                        &                                           & 6000                                                                            & 4                                 \\
4.69061195668268E+018                    & 16.82167                                                                        & -71.35748                                                                       & 804.726807                                                                            & 17.588                                 & 0.013                                     & 18.738                                 & 0.02                                      & 17.34                                  & 0.022                                     &                                        &                                           &                                        &                                           &                                        &                                           & 19.27                                  & 0.066                                     & 5900                                                                            & 4                                 \\
4.69061553009542E+018                    & 17.19522                                                                        & -71.29023                                                                       & 2148.69604                                                                            & 21.492                                 & 0.085                                     &                                        &                                           & 20.997                                 & 0.163                                     &                                        &                                           &                                        &                                           &                                        &                                           &                                        &                                           & 6200                                                                            & 5                                 \\
4.69062494466375E+018                    & 16.66549                                                                        & -71.29562                                                                       & 1007.41357                                                                            & 18.421                                 & 0.027                                     & 19.564                                 & 0.04                                      & 18.335                                 & 0.048                                     &                                        &                                           &                                        &                                           &                                        &                                           & 19.927                                 & 0.09                                      & 6200                                                                            & 5                                 \\
4.69062494466376E+018                    & 16.65732                                                                        & -71.30435                                                                       & 98.0849609                                                                            & 14.948                                 & 0.005                                     & 14.361                                 & 0.005                                     & 13.126                                 & 0.005                                     & 17.547                                 & 0.028                                     & 18.679                                 & 0.036                                     & 19.115                                 & 0.045                                     & 15                                     & 0.009                                     & 6000                                                                            & 4                                 \\
4.6906144649436E+018                     & 16.89218                                                                        & -71.40272                                                                       & 1923.59192                                                                            & 19.756                                 & 0.029                                     & 21.188                                 & 0.05                                      & 19.706                                 & 0.053                                     &                                        &                                           &                                        &                                           &                                        &                                           &                                        &                                           & 5800                                                                            & 4                                 \\
4.69063133557504E+018                    & 16.77777                                                                        & -71.22784                                                                       & 5439.14746                                                                            & 19.785                                 & 0.051                                     & 20.799                                 & 0.071                                     & 19.576                                 & 0.085                                     &                                        &                                           &                                        &                                           &                                        &                                           &                                        &                                           & 5900                                                                            & 5                                 \\
4.69071101149303E+018                    & 17.82407                                                                        & -71.1799                                                                        & 2426.38867                                                                            & 20.819                                 & 0.083                                     & 21.787                                 & 0.112                                     & 20.582                                 & 0.135                                     &                                        &                                           &                                        &                                           &                                        &                                           &                                        &                                           & 6100                                                                            & 4                                 \\
4.69061717936283E+018                    & 17.34527                                                                        & -71.24876                                                                       & 2390.42529                                                                            & 19.987                                 & 0.033                                     & 21.234                                 & 0.053                                     & 19.843                                 & 0.056                                     &                                        &                                           &                                        &                                           &                                        &                                           &                                        &                                           & 6100                                                                            & 4                                 \\
4.69071015679753E+018                    & 17.70098                                                                        & -71.2569                                                                        & 594.829712                                                                            & 19.153                                 & 0.022                                     & 20.514                                 & 0.036                                     & 18.713                                 & 0.033                                     &                                        &                                           &                                        &                                           &                                        &                                           &                                        &                                           & 5500                                                                            & 4.5                               \\
4.69055144917987E+018                    & 17.7799                                                                         & -71.50898                                                                       & 764.347961                                                                            & 17.901                                 & 0.022                                     & 19.022                                 & 0.031                                     & 17.65                                  & 0.036                                     &                                        &                                           &                                        &                                           &                                        &                                           & 19.331                                 & 0.076                                     & 6100                                                                            & 5                                 \\
4.69064449533761E+018                    & 17.88163                                                                        & -71.51531                                                                       & 1103.61462                                                                            & 19.682                                 & 0.049                                     & 21.184                                 & 0.085                                     & 19.374                                 & 0.079                                     &                                        &                                           &                                        &                                           &                                        &                                           &                                        &                                           & 5800                                                                            & 4.5                               \\
4.69061951582505E+018                    & 17.0205                                                                         & -71.24934                                                                       & 1337.36548                                                                            & 17.618                                 & 0.013                                     & 18.464                                 & 0.017                                     & 17.466                                 & 0.023                                     & 21.768                                 & 0.144                                     & 21.95                                  & 0.178                                     & 22.245                                 & 0.191                                     & 18.612                                 & 0.035                                     & 6600                                                                            & 5                                 \\
4.6905509681436E+018                     & 17.49441                                                                        & -71.5691                                                                        & 1432.04077                                                                            & 20.309                                 & 0.066                                     &                                        &                                           & 19.894                                 & 0.101                                     &                                        &                                           &                                        &                                           &                                        &                                           &                                        &                                           & 5900                                                                            & 5                                 \\
4.69062043923299E+018                    & 17.29293                                                                        & -71.1                                                                           & 1625.68982                                                                            & 20.146                                 & 0.043                                     & 21.416                                 & 0.071                                     & 19.937                                 & 0.073                                     &                                        &                                           &                                        &                                           &                                        &                                           &                                        &                                           & 6200                                                                            & 4.5                               \\
4.69061298747482E+018                    & 16.75535                                                                        & -71.32757                                                                       & 1448.65942                                                                            & 21.356                                 & 0.105                                     &                                        &                                           &                                        &                                           &                                        &                                           &                                        &                                           &                                        &                                           &                                        &                                           & 5600                                                                            & 4                                 \\
4.6906198594224E+018                     & 17.13099                                                                        & -71.20618                                                                       & 2383.63672                                                                            & 19.409                                 & 0.03                                      & 20.315                                 & 0.041                                     & 19.249                                 & 0.052                                     &                                        &                                           &                                        &                                           &                                        &                                           & 20.639                                 & 0.091                                     & 6400                                                                            & 5                                 \\
4.69065947618345E+018                    & 18.01465                                                                        & -71.29682                                                                       & 1934.09143                                                                            & 20.485                                 & 0.071                                     &                                        &                                           & 20.013                                 & 0.104                                     &                                        &                                           &                                        &                                           &                                        &                                           &                                        &                                           & 5800                                                                            & 4                                 \\
4.69056220377801E+018                    & 17.18624                                                                        & -71.52321                                                                       & 2001.50391                                                                            & 20.222                                 & 0.063                                     & 21.349                                 & 0.092                                     & 20.169                                 & 0.114                                     &                                        &                                           &                                        &                                           &                                        &                                           &                                        &                                           & 6100                                                                            & 3.5                               \\
4.69056584591023E+018                    & 17.25588                                                                        & -71.46314                                                                       & 392.160675                                                                            & 19.036                                 & 0.021                                     & 20.413                                 & 0.035                                     & 18.464                                 & 0.03                                      &                                        &                                           &                                        &                                           &                                        &                                           &                                        &                                           & 5300                                                                            & 4.5                               \\
4.69061652222265E+018                    & 17.55176                                                                        & -71.26899                                                                       & 3659.71973                                                                            & 20.987                                 & 0.052                                     &                                        &                                           & 20.902                                 & 0.113                                     &                                        &                                           &                                        &                                           &                                        &                                           &                                        &                                           & 6600                                                                            & 4.5                               \\
4.6907148297219E+018                     & 17.61696                                                                        & -71.13598                                                                       & 3342.55542                                                                            & 20.816                                 & 0.083                                     &                                        &                                           & 20.594                                 & 0.135                                     &                                        &                                           &                                        &                                           &                                        &                                           &                                        &                                           & 6300                                                                            & 4                                 \\
4.69062119945219E+018                    & 17.28052                                                                        & -71.1939                                                                        & 8176.15332                                                                            & 20.519                                 & 0.052                                     & 21.951                                 & 0.121                                     & 20.186                                 & 0.081                                     &                                        &                                           &                                        &                                           &                                        &                                           &                                        &                                           & 5400                                                                            & 5                                 \\
4.69065782691309E+018                    & 17.87784                                                                        & -71.40297                                                                       & 1653.60144                                                                            & 18.981                                 & 0.025                                     & 19.933                                 & 0.034                                     & 18.911                                 & 0.045                                     &                                        &                                           &                                        &                                           &                                        &                                           & 20.16                                  & 0.111                                     & 6400                                                                            & 4.5                               \\
4.69056670490367E+018                    & 17.16157                                                                        & -71.43209                                                                       & 208.265167                                                                            & 21.065                                 & 0.054                                     &                                        &                                           & 19.962                                 & 0.062                                     &                                        &                                           &                                        &                                           &                                        &                                           &                                        &                                           & 4600                                                                            & 5.5                               \\
4.69063356895797E+018                    & 17.18656                                                                        & -71.12657                                                                       & 2213.52783                                                                            & 19.39                                  & 0.042                                     & 20.407                                 & 0.059                                     & 19.238                                 & 0.073                                     &                                        &                                           &                                        &                                           &                                        &                                           &                                        &                                           & 6400                                                                            & 4                                 \\
4.69056732337895E+018                    & 17.46448                                                                        & -71.43648                                                                       & 165.811844                                                                            &                                        &                                           &                                        &                                           & 21.177                                 & 0.113                                     &                                        &                                           &                                        &                                           &                                        &                                           &                                        &                                           & 3900                                                                            & 6                                 \\
4.69055117430198E+018                    & 17.58524                                                                        & -71.52973                                                                       & 508.489746                                                                            & 21.257                                 & 0.101                                     &                                        &                                           & 20.537                                 & 0.136                                     &                                        &                                           &                                        &                                           &                                        &                                           &                                        &                                           & 5000                                                                            & 5.5                               \\
4.69056467767907E+018                    & 17.74985                                                                        & -71.40001                                                                       & 643.350525                                                                            & 20.903                                 & 0.061                                     &                                        &                                           & 20.098                                 & 0.078                                     &                                        &                                           &                                        &                                           &                                        &                                           &                                        &                                           & 5200                                                                            & 4                                 \\
4.69061968762373E+018                    & 17.00041                                                                        & -71.2434                                                                        & 1048.64355                                                                            & 21.527                                 & 0.082                                     &                                        &                                           & 20.904                                 & 0.12                                      &                                        &                                           &                                        &                                           &                                        &                                           &                                        &                                           & 5400                                                                            & 5                                 \\
4.69056213505854E+018                    & 17.1882                                                                         & -71.53268                                                                       & 3556.75244                                                                            & 20.643                                 & 0.076                                     & 21.403                                 & 0.094                                     & 20.525                                 & 0.135                                     &                                        &                                           &                                        &                                           &                                        &                                           &                                        &                                           & 6500                                                                            & 4                                 \\
4.69062016866005E+018                    & 17.31147                                                                        & -71.22773                                                                       & 3898.67773                                                                            & 21.548                                 & 0.116                                     &                                        &                                           &                                        &                                           &                                        &                                           &                                        &                                           &                                        &                                           &                                        &                                           & 5600                                                                            & 5                                 \\
4.69061487726044E+018                    & 16.98475                                                                        & -71.35109                                                                       & 1013.1073                                                                             & 18.733                                 & 0.018                                     & 20.072                                 & 0.03                                      & 18.631                                 & 0.032                                     &                                        &                                           &                                        &                                           &                                        &                                           & 20.661                                 & 0.081                                     & 5800                                                                            & 4                                 \\
4.69061916792237E+018                    & 16.86111                                                                        & -71.27514                                                                       & 2806.41772                                                                            & 21.509                                 & 0.083                                     &                                        &                                           & 21.154                                 & 0.175                                     &                                        &                                           &                                        &                                           &                                        &                                           &                                        &                                           & 7000                                                                            & 3                                 \\
4.69061965326398E+018                    & 17.10539                                                                        & -71.22253                                                                       & 1057.13074                                                                            & 19.754                                 & 0.036                                     & 21.153                                 & 0.065                                     & 19.781                                 & 0.074                                     &                                        &                                           &                                        &                                           &                                        &                                           &                                        &                                           & 5900                                                                            & 3.5                               \\ \bottomrule
\end{tabular}%
}
\begin{tablenotes}
    \item \textbf{Note:} This table is truncated to the first 50 entries. The entire contents of this table are available in the electronic version of the article.
           
\end{tablenotes}
\end{table}
\end{landscape}

\end{appendix}

\end{document}